%% file: main.tex
\documentclass[acmsmall]{acmart}

\usepackage{amsmath,amsfonts}
\usepackage{algorithmic}
\usepackage{graphicx}
\usepackage{textcomp}
\usepackage{xcolor}
\usepackage{booktabs}
\usepackage{color, colortbl}
\definecolor{Gray}{gray}{0.9}

\usepackage{multirow}
\usepackage[framemethod=TikZ]{mdframed}
\usepackage{xcolor}
\usepackage{tcolorbox}[boxsep=1mm]

\usepackage[T1]{fontenc}
\usepackage[]{babel}
\usepackage[utf8]{inputenc}

\usepackage{longtable}
\usepackage{lscape}
\usepackage{afterpage}

\newcommand{\news}[1]{\noindent\textbf{#1:}}

\mdfdefinestyle{implicationBox}{
    linecolor=blue,
    outerlinewidth=2pt,
    roundcorner=10pt,
    innertopmargin=10pt,
    innerbottommargin=10pt,
    innerrightmargin=10pt,
    innerleftmargin=10pt,
    backgroundcolor=gray!10!white
}

\UseRawInputEncoding

\AtBeginDocument{%
  \providecommand\BibTeX{{%
    \normalfont B\kern-0.5em{\scshape i\kern-0.25em b}\kern-0.8em\TeX}}}

\setcopyright{acmcopyright}
\copyrightyear{2018}
\acmYear{2018}
\acmDOI{XXXXXXX.XXXXXXX}

\acmJournal{TOSEM}
\acmVolume{37}
\acmNumber{4}
\acmArticle{111}
\acmMonth{8}

\begin{document}

\title{Accountability in Code Review: The Role of Intrinsic Drivers and the Impact of LLMs}

\author{Adam Alami}
\email{adal@mmmi.sdu.dk}

\affiliation{%
  \institution{The University of Southern Denmark}
  \streetaddress{The Maersk Mc-Kinney Moller Institute}
  \city{S\o nderborg}
  \country{Denmark}
  \postcode{6400}
}

\author{Victor Vadmand Jensen}
\affiliation{%
  \institution{Aarhus University}
  \streetaddress{Jens Chr. Skous Vej 4}
  \city{Aarhus C}
  \country{Denmark}}
  \postcode{8000}
\email{vvj@clin.au.dk}

\author{Neil A. Ernst}
\affiliation{%
  \institution{University of Victoria}
  \streetaddress{PO Box 1700 STN CSC}
  \city{Victoria BC V8W 2Y2}
  \country{Canada}}
\email{nernst@uvic.ca}

\renewcommand{\shortauthors}{Alami, et al.}

\begin{abstract}

Accountability is an innate part of social systems. It maintains stability and ensures positive pressure on individuals' decision-making. As actors in a social system, software developers are accountable to their team and organization for their decisions. However, the drivers of accountability and how it changes behavior in software development are less understood. In this study, we look at how the social aspects of code review affect software engineers' sense of accountability for code quality. Since software engineering (SE) is increasingly involving Large Language Models (LLM) assistance, we also evaluate the impact on accountability when introducing LLM-assisted code reviews. We carried out a two-phased sequential qualitative study ($\textbf{interviews} \rightarrow \textbf{focus groups}$). In Phase I (16 interviews), we sought to investigate the intrinsic drivers of software engineers influencing their sense of accountability for code quality, relying on self-reported claims. In Phase II, we tested these traits in a more natural setting by simulating traditional peer-led reviews with focus groups and then LLM-assisted review sessions. We found that there are four key intrinsic drivers of accountability for code quality: \emph{personal standards}, \emph{professional integrity}, \emph{pride in code quality}, and \emph{maintaining one's reputation}. In a traditional peer-led review, we observed a transition from \emph{individual} to \emph{collective accountability} when code reviews are initiated. We also found that the introduction of LLM-assisted reviews disrupts this accountability process, challenging the reciprocity of accountability taking place in peer-led evaluations, i.e., one cannot be accountable to an LLM. Our findings imply that the introduction of AI into SE must preserve social integrity and collective accountability mechanisms.

\end{abstract}

\ccsdesc[500]{Software and its engineering~Programming teams}

\keywords{Code quality, Accountability, Artificial Intelligence, Large Language Models, LLM, Code Review, Human and Social Aspects of Software Engineering}

\received{20 February 2024}
\received[revised]{12 March 2025}
\received[accepted]{5 June 2025}

\maketitle

\input{introduction.tex}

\input{theory}
\input{related.tex}

\input{methods.tex}
\input{findings.tex}

\input{discussion.tex}
\input{validity.tex}

\input{conclusion.tex}

\begin{acks}

We would like to thank our interviewees and the focus groups participants for their time and effort in making this study possible. This work was funded by the first author's tenure-track position funding available for research and provided by the Computer Science department at Aalborg University.

\end{acks}

\bibliographystyle{ACM-Reference-Format}
\bibliography{references}

\end{document}

%% file: introduction.tex
\section{Introduction}\label{sec:introduction}

\noindent \textbf{Accountability is the expectation of being held responsible for one's actions and the need to provide explanations and reasoning for such actions to others in the future} \citep{de2007justifying,lerner1999accounting,frink1998toward}. To software engineers, accountability for code quality is to adhere and meet expectations and justify their coding decisions. Either these expectations are set at the organizational or team levels, or they align with widely recognized quality standards~\citep{alami2024understanding}.

Developing software is done as part of a social system. Accountability maintains stability in these social systems~\citep{frink2004advancing}. Accountability mechanisms thus have profound consequences for core organizational functions, yet the software engineering (SE) research community has paid little attention to them \citep{alami2024understanding}. The absence of accountability in a social system like software development may result in individuals acting with little regard to the consequences imposed by others \citep{hall2017accountability}. Consequently, organizations may find it challenging to effectively manage their operations \citep{frink1998toward}. 

A key area of accountability in software engineering is deciding which changes to make in a codebase and the applicable quality standards. These decisions are often taken as part of a socio-technical process called code review \citep{meneely2014empirical,bosu2016process}. Code reviews can lead to \emph{social dilemmas} because individual and collective interests may conflict \citep{de2009paying,de2001less}. A social dilemma is a situation when individuals are faced with a choice between pursuing their own interests at the expense of a group or cooperating for the greater benefit of the collective \citep{de2009paying,de2001less}. The dilemma arises because if everyone pursues their own self-interest, the collective outcome is worse than if everyone cooperated \citep{de2009paying,de2001less}.

In our previous work, we reported that intrinsic drivers such as personal standards and professional integrity underpin software engineers' sense of accountability for code quality \citep{alami2024understanding}. They shape how developers perceive and respond to their responsibilities in meeting code quality expectations. Peer-led review is a well-established code quality assurance practice, where peers review each other's code for errors, readability, and adherence to agreed coding standards \citep{mcintosh2016empirical,alami2019affiliated}. Group dynamics, human interactions, and mutual expectations influence accountability in this process, bringing a social dimension to it \citep{alami2024understanding}.

Large Language Models (LLMs) like GPT-4 are powerful multipurpose products, capable of generating code and assisting in software development activities, including documentation, debugging, design suggestions, and code review \cite{fan2023large}. The versatility and broad applicability of LLMs have the potential to influence the landscape of software engineering by either reducing the reliance on human-led activities or supporting tasks that were previously considered too complex for automation, including code review \cite{svyatkovskiy2020,zohair2018future}.
 
However, SE activities and practices are highly collaborative. The introduction of LLM-assisted reviews to traditionally human-led activities may influence or even shift accountability dynamics. The reduction in human interactions may weaken the reinforcement of team norms and informal accountability mechanisms used to enforce it \citep{alami2024understanding}.

During code reviews, software developers may face pressure to justify their decisions and actions to the rest of the group, fearing negative evaluations, a type of social dilemma known as a public good dilemma~\citep{de2008reputational,leary2019self,tyler1999people}. They need to justify their coding decisions, and failing to do so may tarnish their reputation as reliable and cooperative team members. 

Analogous to social dilemmas, where individuals are evaluated against group expectations \cite{de2007justifying}, software developers in code reviews face similar conditions. They must navigate justifying and defending their coding decisions and maintaining their image as competent and collaborative team members. Drawing from the broader social dilemma literature \cite{de2003accountability,de2001less}, where the failure to align with group expectations risks one being perceived as untrustworthy, non-cooperative, and sometimes leading to reputational consequences \cite{de2009paying,de2007justifying}. Social dilemma research has shown that individuals exhibit reduced self-interest when they are subjected to accountability measures \cite{de2009paying,de2007justifying}. To understand whether the condition of being held accountable in code review influences developers' code quality, our study aims to explore the intrinsic drivers that motivate software engineers in these contexts and how they influence their sense of accountability. By examining the impact of both peer-led and LLM-assisted code reviews, we seek to learn the underlying mechanisms that shape accountability for code quality within the socio-technical system of software development. 

Human and social aspects of code review have received little attention from SE research~\citep{davila2021systematic,badampudi2023modern}. We elaborate on the few studies that have, later in Sect. \ref{sec:related}. Like most SE research, code review studies have primarily focused on technical artifact analysis in open source communities~\citep{Storey2020}. 

Building on our earlier study that laid a basic understanding of accountability in SE \citep{alami2024understanding}, we now delve into software engineers' innate qualities, which we term their \emph{intrinsic drivers} \cite{deci2000and}. Recognizing the importance of social dilemmas within the code review process, we seek to understand software engineers' intrinsic drivers, how they affect accountability during peer-led code reviews, and the impact of the integration of LLM-assisted code reviews. With this focus, we aim to provide an understanding of accountability for code quality decisions within the socio-technical system of software development, beyond the technical aspects. 

\textbf{Intrinsic drivers} are factors such as adhering to self-imposed personal standards and reputation that drive behavior and influence outcomes \citep{cameron1994reinforcement,fishbach2022structure}. Our previous work showed the importance of intrinsic drivers for software engineers, nurturing their feeling of accountability for their work \citep{alami2024understanding}. However, how these intrinsic qualities shape individuals' accountability is unreported in both the accountability theories \citep{frink1998toward,frink2004advancing} and software engineering \citep{alami2024understanding}, leaving a gap in understanding the role of intrinsic drivers in influencing accountability in SE. Therefore, we propose exploring intrinsic drivers within SE to fully comprehend their impact on professional behavior and outcomes.

Understanding intrinsic drivers' role in driving accountability for code quality is a relevant software engineering problem. Quality assurance practices like testing and code review serve as accountability mechanisms for ensuring code quality \citep{alami2024understanding,bosu2013impact}. Without a clear understanding of how the human and social factors, such as intrinsic drivers influence accountability, organizations risk relying on extrinsic motivators (e.g., promotion) and technical processes (e.g., continuous integration), which alone may not be as effective in sustaining long-term accountability as relying on both personal qualities and other social enablers.

In addition, the increased adoption of artificial intelligence (AI) in SE \citep{fan2023large} introduces a potential alteration to established social dynamics due to reduced opportunities for peer interaction and group norms. Therefore, for designing interventions that preserve the social fabric of software engineering teams in future AI-augmented SE, it is crucial to understand the influence of intrinsic drivers in both peer-led and LLM-assisted contexts. For example, engineers who derive accountability from professional integrity and peer validation may struggle to adapt in an environment where feedback is predominantly AI-driven. This understanding will also lay the foundation for future work to create more resilient accountability practices that integrate human and AI collaboration effectively.

In our previous work, we identified several outcomes software engineers bear accountability for, namely meeting deadlines, software security, and code quality \citep{alami2024understanding}. We also examined a broad set of drivers, including extrinsic motivators such as financial rewards. 

While our previous work focused on the building blocks of accountability \citep{alami2024understanding}, in this study, we aim to focus on a specific SE outcome, code quality, in a common SE practice, code review. We build on our prior findings by investigating how intrinsic drivers influence accountability in a specific SE practice and how an evolving technology, such as LLMs, might reshape the evolution of accountability at the individual and team levels. By exploring accountability within a socio-technical context like code review, we aim to identify the interplay between human drivers, accountability for an important SE outcome, and potential implications of AI integration. By doing so, we seek to address gaps left unexamined in our earlier work.

In this study, we expand on this work and focus on a particular outcome, code quality, a critically important SE outcome \citep{alami2022scrum,vasilescu2015quality}. This leads us to propose:

\medskip
    
    \noindent \textbf{RQ1:} What are the key intrinsic drivers of software engineers that influence their sense of accountability towards the quality of their code?
    
\medskip

To address \textbf{RQ1}, we carried out a two-phased sequential qualitative study ($\textbf{interviews} \rightarrow \textbf{focus groups}$) \citep{creswell2017designing}. In Phase I (16 interviews), we sought to investigate the intrinsic drivers of software engineers influencing their sense of accountability for code quality, relying on self-reported claims. However, relying on self-reported data for complex and subjective topics like accountability is subject to biases and inaccuracies caused by social desirability bias, recall bias, and self-perception \citep{podsakoff2003common}. Therefore, in Phase II, we tested these traits and claims in a more natural setting by simulating traditional peer-led reviews using four focus groups (5--6 participants each). This design allowed us to compensate for the inherent methodological weaknesses of each phase, a strength of mixed methods studies \citep{creswell2017designing}. By design, focus groups foster social interaction~\citep{Kontio2008}, which can test self-reported intrinsic drivers influencing personal accountability in professional-like settings~\citep{carroll2003making}. This interaction helped us to further understand how software engineers navigate the challenges of maintaining their intrinsic drivers for accountability in a complex social setting.

In Phase I of the study, we identified four intrinsic drivers that influence software engineers' sense of accountability to meet established expectations for code quality: \emph{personal standards}, \emph{professional integrity}, \emph{pride in code quality}, and \emph{maintaining personal reputation}.

In Phase I and our previous work \citep{alami2024understanding}, code review emerged as a significant accountability mechanism for code quality. Participants frequently referred to code review as a central process to demonstrate and enforce accountability. This emergent finding provided a natural basis for shifting the focus of the secondary analysis to code review, aligning with the broader goal of understanding accountability in socio-technical systems. Moreover, participants emphasized the relevance of code review as an accountability mechanism. These discussions underscored its importance as both a collaborative and evaluative process, making it a relevant context for deeper exploration in the secondary analysis.

Recent advances in AI technologies such as neural network architectures, recurrent neural networks, and transformers \cite{vaswani2017attention} have led to the development of several commercial products such as Tabnine\footnote{\url{https://www.tabnine.com/}}, CodeX\footnote{\url{https://openai.com/index/openai-codex/}}, and Github's Copilot\footnote{\url{https://github.com/features/copilot/}}. Although these products are mainly directed to code generation, Large Language Models (LLM) are multi-task products capable of assisting in other SE tasks \citep{fan2023large,lu2023llama}, traditionally governed by humans.

As efforts in code review automation are growing, e.g., \cite{pandya2022corms,shi2019automatic}, the \textbf{integration of LLMs} into code review is either already happening \cite{lu2023llama,li2022automating} or just a matter of time \cite{lu2023llama}. This may bring a significant shift from traditional code review. For example, in response to \textbf{RQ1}, we learned that engineers feel accountable for the quality of their code to their peers. In an LLM-assisted code review, by comparison, human feedback and interaction are replaced by a machine equivalent. This shift has prompted us to examine the resilience and perseverance of human factors, such as the desire to maintain professional integrity, in a machine-based evaluation process. 

While \textbf{RQ1} delves into the intrinsic traits driving software engineers' accountability in a human-led code review process, \textbf{RQ2} seeks to understand the ramifications for accountability in LLM-assisted code review. Most SE activities are highly collaborative. Thus, the integration of AI into the already complex human and social practices requires a thorough understanding of potential implications. Specifically, the introduction of AI in practices traditionally dominated by human expertise raises questions about the human-AI engagement and behavioral responses of software engineers. For example, Malone et al. suggest that the most challenging changes in AI integration are not computers replacing humans but rather people and computers working together as an integrated approach \cite{malone2020artificial}. To effectively augment SE practices with the integration of AI, it is crucial to understand how software engineering teams engage with and respond to AI tools in their workflows. AI is a newcomer to an already socially loaded process like code review, where accountability for code quality---a key and highly sought-after SE outcome \citep{vasilescu2015quality,krasner2018,alami2022scrum}---is shaped by complex human interactions and social norms. AI integration into SE raises many questions about how it may influence the social fabric of these processes, especially in terms of accountability. Therefore, we inquire:

\medskip

    \noindent \textbf{RQ2:} How does LLM-assisted code review affect software engineers' accountability for code quality?
    
\medskip

LLM-assisted code review refers to the process of utilizing a Large Language Model, such as GPT-4 or Gemini, to evaluate and provide feedback on software code. For our study, we designed a specific prompt for the LLM to simulate the role of a reviewer by analyzing the attached code and generating feedback (see Sect. \ref{sec:methods}).

To address \textbf{RQ2}, we carried out LLM-assisted reviews in the second part of our focus groups. Our focus groups were scheduled over two hours. In the first hour, we conducted peer-led reviews (\textbf{RQ1}), and in the second hour, LLM-assisted reviews (further details in Sect. \ref{sec:methods}). 

We contribute:

\begin{itemize}

    \item [-] \textbf{Elucidation of individual and collective accountability in code review:} Our study provides a comprehensive examination of how intrinsic drivers such as professional integrity, pride, personal standards, and reputation influence software engineers' sense of individual accountability towards code quality. During the code review, we identified a complex accountability process. Accountability for code quality shifts from the individual-level when writing code to the collective-level when peers reciprocate accountability for code quality during the review. This contribution shows that code quality is beyond standards, processes, or metrics but a shared value that is cultivated through a sense of individual and collective accountability. This implies that code quality is not solely achieved through tools and technical practices like testing and coding standards but is also deeply personal and collectively fostered through social norms and human interactions. Our findings underscores the importance of balancing the ``socio'' and ``technical'' dimensions in pursuing code quality. By integrating both human-driven and technical practices, SE teams can create environments where accountability for code quality is cultivated through shared values.

    \item [-] \textbf{Insights into the impact of LLM-assisted code review:} Our work assesses the impact of LLM-assisted code review on accountability for code quality. We provide insights into the challenges and disruptions introduced by integrating AI technologies into traditional SE practices. When we introduced LLM-assisted reviews, we learned that this newcomer disrupts the collective sense of accountability for code quality. This disruption does not only accentuate the social fundamentals of code review but also emphasizes the need for designing AI integration in SE that supports rather than disrupts the social fabric of SE teams. While AI will continue augmenting SE practices, our findings inform future research and organizations that integrating AI into the inherently socio-technical processes of SE is not a simple out-of-the-box integration. The disruption to collective accountability demonstrates the importance of considering the social impact of AI integration. The design of future AI integration in SE should account for the social fabric of SE teams.

    \item [-] \textbf{Identification of LLM-related factors disrupting collective accountability for code quality:} We identified LLM-related factors that contribute to the disruption of collective accountability within code review, mainly trust in LLMs and their inherent limitations. This contribution emphasizes the importance of addressing trust in AI technologies and preserving the social aspects of the adoption process. Our findings show that software engineers remain skeptical about the reliability of AI as a partner. This trust issue stems from the training constraints of the AI tools and limitations in integrating seamlessly into SE processes. These findings highlight that the technology's current shortcomings undermine confidence in its ability to support SE effectively. Addressing these limitations is essential for advancing AI tools as reliable partners in SE processes.
    
\end{itemize}

We structure the remainder of this paper as follows: In the next section \ref{sec:theory}, we discuss the theoretical frameworks reported in social and organizational sciences literature. In Sect. \ref{sec:related}, we discuss related work. Section \ref{sec:methods} describes our research design and methods. Section \ref{sec:findings} is dedicated to the data interpretation from our first and second phases of the study. The implications of our results, in both existing theory and practice, are then discussed in Sect. \ref{sec:discussion}. In section \ref{sec:trust}, we discuss the study's trustworthiness, limitations, and trade-offs. Section \ref{sec:conclusion} brings us to a close.

%% file: theory.tex
\section{Theoretical Background}\label{sec:theory}

\noindent Accountability is considered the link between individuals and their social system, creating an identity relationship associating individuals with their actions and performance \citep{tetlock1992impact,mero2007accountability}. Accountability has implications in all organizational levels \citep{frink1998toward}. However, our focus, in this study, is the micro-level or \emph{felt accountability} (referred to simply as ``accountability'') which refers to the individual's perception of accountability \citep{frink1998toward}; the most studied type of accountability \citep{hall2017accountability}. This type of accountability is contingent upon the individual's own interpretation \citep{frink1998toward}, rather than the external milieu imposed expectations \citep{folger2001fairness}.

Irrespective of the nature of a collective, ranging from a dyad to a civilization, addressing coordination and collaboration among its constituent members with diverse interests is imperative \citep{schlenker1994triangle}. Accountability comes to play by establishing shared expectations in social systems. When standards are set, people must adhere to them and the failure to do so may result in imposing penalties \citep{schlenker1994triangle,frink1998toward}. Compliance with standards is evaluated at various layers of social systems, including the individual, the dyad, the group, the organization, and the society as a whole \citep{gelfand2004culture,frink1998toward}.

The concept of accountability inherently implies an anticipated evaluation \citep{hall2017accountability}. For the latter to take place, the individual must engage in an account-giving process \citep{frink2008meso}, which may results in rewards or sanctions, and its legitimacy is affirmed by an audience \citep{hall2017accountability}. The role of the audience is to evaluate performance using rules, standards, and expectations and distribute rewards or recommend punishments based on the outcome of the evaluations \citep{hall2017accountability}. Schlenker et al. explained that ``accounts'' shared in the process are fundamental to the process as they either enhance, protect, or damage the individuals' self-image \citep{schlenker1994triangle}. Tetlock argued that individuals often ``defend'' their actions when facing evaluations to protect their self-image and status, highlighting accountability as a fundamental social contingency driving individual behavior and decisions \citep{tetlock1992impact}.

Drawing from several conceptualization frameworks \citep{frink1998toward,schlenker1994triangle,tetlock1992impact}, the integration of accountability in a social systems aims at creating social order \citep{hall2004leader}. Social systems employ accountability mechanisms to cultivate structured environments and social order, where individuals are expected to be held accountable for their participation in various social activities \citep{hall2017accountability}.

When faced with accountability demands, individuals develop coping mechanisms, both proactive and reactive, to maintain a consistent image of themselves \citep{schlenker1994triangle,hall2017accountability}. Finally, given its implications, individuals resort to avoiding, manipulating, or otherwise coping with their accountabilities \citep{tetlock1992impact,hall2017accountability}. For example, Tetlock explained that individuals engage in cognitive laziness, adjusting their account-giving in advance to align with the audience preferences or using their most easily defensible options (acceptability heuristic) to explain their actions \citep{tetlock1992impact}. However, if they discover that they are accountable after the actions have occurred, then, they may engage in retrospective rationality, defending their past behaviors with justifications and excuses \citep{lerner1999accounting}.

In this intricate context, Frink and Klimoski suggest that any conceptualization of workplace accountability should consider both the formal and the informal manifestation of accountability \citep{frink1998toward}. While informal accountability is formalized over time and grounded in organizational rules and policies \citep{frink2008meso}, informal (also referred to as accountability for others) is the perceived accountability ``outside of or beyond formal position or organizational policy'' \citep{zellars2011accountability}. Informal accountability is grounded in unofficial expectations and discretionary behaviors that result from the socialization of network members \citep{romzek2012preliminary}. Shared norms also lay out an informal code of conduct used by group members as a reference for appropriate and inappropriate behaviors \citep{romzek2012preliminary}. Romzek et al. found that informal accountability in nonprofit networks is fostered by trust, reciprocity, and respect for institutional turf \citep{romzek2012preliminary}. Similarly, informal accountability is exercised through evaluations that result in either rewards or sanctions \citep{romzek2014informal}, but remain informal in nature. For example, rewards can be in the form of favors and public recognition, and sanctions may lead to reduced reputation, loss of opportunities within the group, and exclusion from future information sharing \citep{romzek2014informal}.

\begin{figure*}[t!]

    \centering
    \includegraphics[ trim=2cm 3cm 1cm 2cm, clip,
      width = 1.0 \textwidth
    ]{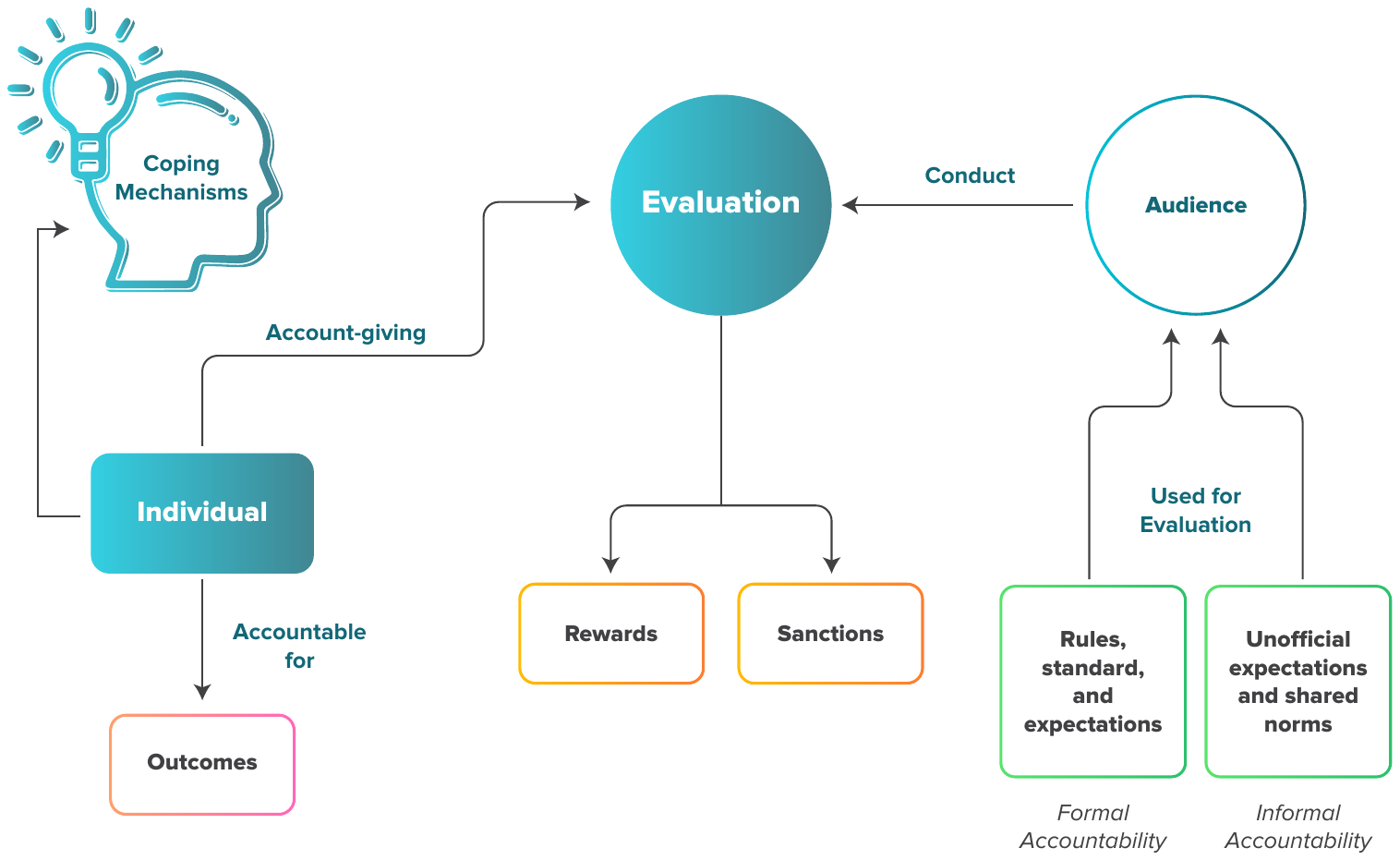}

    \caption{The Dynamics of Accountability in Social Systems}%
    \label{fig:figure_1}
\Description[]{Map of theory of how accountability works in social systems. Audiences evaluate individuals, using rewards and sanctions.}
\end{figure*}

Figure \ref{fig:figure_1} is a visual representation of the theorization of accountability. In sum, accountability is rooted in social systems and the individuals are at its core. It consists of an evaluation process where individuals are held answerable for their actions and decisions, guided by interpersonal, social, and structural factors within specific sociocultural contexts. Performance against predefined rules, standards, official and unofficial expectations, and shared norms is assessed by an audience resulting in the dispensation of rewards or sanctions. In response, individuals develop various coping mechanisms, both proactive and reactive, to safeguard and maintain a consistent self-image within their social system. Some of these mechanisms may include cognitive strategies as aligning actions with audience preferences or retrospective rationalization, where individuals defend past behaviors with justifications and excuses. While traditional accountability theory has often treated the ``audience'' as a singular entity, Gelfand et al. argue that it can be deconstructed into a web of multiple parties to whom the individual is answerable \citep{gelfand2004culture}.

\subsection{Operationalization of accountability}

In software engineering, formal accountability is operationalized and controlled through institutionalized mechanisms that are explicitly designed to enforce it, such as performance evaluations, established coding standards, and formal performance reviews. These mechanisms aim to align individual behavior with organizational goals by providing clear expectations and consequences, whether through rewards such as promotions or punishments like performance improvement plans. For instance, performance metrics might include defect rates in code or adherence to project deadlines. These mechanisms aim to ensure that accountability is tied to the outcomes set by the organization \citep{alami2024understanding}. In our previous work, we found that software engineers are being individually held accountable for code quality, software security, and meeting project deadlines \citep{alami2024understanding}.

In contrast, informal accountability operates through the shared norms and expectations within teams. Rooted in interpersonal relationships and mutual trust, it is reinforced through peer feedback and team interactions rather than formal rules. As highlighted in our previous research \citep{alami2024understanding}, informal accountability is cultivated through intrinsic motivations and social drivers such as meeting peers expectations, reciprocity of accountability, and adhering to the team's standards. For example, a software engineer may strive for high code quality to uphold their reputation within the team or to meet peer expectations during code reviews. Unlike formal accountability, which relies on structured and institutionalized evaluations, informal accountability relies on psychologically safe environments where individuals are encouraged to take ownership of their contributions without fear of blame \citep{alami2024understanding}. Both forms of accountability co-exist in software engineering environments, with informal mechanisms often emerging within the team \citep{alami2024understanding}.

%% file: related.tex
\section{Related Work}\label{sec:related}

In this section, we summarize and discuss existing research on human and social factors affecting code review processes. Based on the literature we surveyed, ``social'' aspects pertain to group-level interactions, relationships, and dynamics within the SE environment, such as communication and collaboration. Social also entails behaviors inherent to group dynamics such as trust, perception of fairness, and conflicts. On the other hand, ``human'' aspects are particular to individual' behaviors and characteristics such as cognitive load, motivation, and emotional responses.

\subsection{Social aspects in SE}

In this stream of work, researchers examined the social concerns arising from the inherently human-intensive aspects of the code review. German et al. study fairness by examining how practitioners perceive the treatment they receive from their peers during a code review \citep{german2018my}. They found that a sense of unfairness is predominant among developers. This is more typical with authors of code than reviewers, and there is an often misplaced sense that the merits of the code should dictate acceptance~\citep{german2018my}.

Gon\c{c}alves et al reported that developers find code review conflict-causing \citep{goncalves22}. However, neither that study nor the fairness study \citep{german2018my} examined how conflict or unfairness might impact personal accountability, although Gon\c{c}alves et al noted organizational factors, such as autonomy, might result in developers' improved ``effort in doing a good job.'' Role inequality was, as we will also show, less important as long as the critiques were constructive \citep{goncalves22}. 

Bosu et al. and Bosu and Carver studied reviewers' perceptions of their peers in the review process \citep{bosu2013impact,bosu2016process}, particularly how it emerges and influences the practice. Both studies conclude that the review process influences impression formation, especially peers' competence. Subpar code changes influence perception negatively and subsequently become a perception of authors in future reviews \citep{bosu2013impact,bosu2016process}.

Egelman et al. looked at negative interactions in code review (``pushback'') at Google~\citep{Egelman2020}. They found interpersonal conflict had several causes, such as confrontational comments. 

In our previous work, we found that the process of code review is strategically used as a mechanism for ensuring accountability for code quality, aiming to compel software engineers to adhere to established quality standards \citep{alami2024understanding}.

\subsection{Human Aspects in SE}

This stream of work focuses on specific reviewer behavior \citep{davila2021systematic}. Kitagawa et al. used simulation to understand developer engagement during code review \citep{kitagawa2016code}. They used a situational model based on a snowdrift game. The key finding of the study is that reviewers partake in the review when they perceive that the value of their participation outweighs the cost \citep{kitagawa2016code}. Baum et al. carried out an experiment study to investigate the relationship between reviewers' cognitive burden and their performance during the review \citep{baum2019associating}. The main finding is the relation between working memory capacity and the reviewer's efficacy in detecting delocalized flaws, as well as the negative effect of larger and more complex code changes on reviews' performance \citep{baum2019associating}.

Alami et al. sought to understand the motivation of open-source contributors to participate in code reviews \citep{alami2019does}. They conclude that contributors' motivation is rooted in hackers ethics (e.g., improvement of quality and passion for coding). Despite rejections and toxic communication, contributors show resilience, driven by their intrinsic and extrinsic motives to participate. Some of the intrinsic drivers are the passion for coding and caring about the quality of the code. Extrinsically, contributors partake in code review to showcase their skills and maintain a reputation in the community \citep{alami2019does}.

The distinction between open-source code review and institutional code review is important. Open source code review has been the focus of the majority of the research. However, it is more often \textbf{asynchronous} and \textbf{distributed}, compared to institutional reviews. For example, the systematic review of Badampudi et al.~\citep{badampudi2023modern} explained that ``when discussing reviewer interactions, human aspects received much attention in the reviewed primary studies. However, the investigation of review dynamics, social interactions, and review performance is focused on OSS projects. It is not known if such interactions differ in proprietary projects'' ~\citep{badampudi2023modern}.

\subsection{Accountability in SE}

Our earlier study \citep{alami2024understanding} examined the broader concept of accountability in SE and was not directly concerned with code review. That broad investigation of individual accountability within software engineers highlighted two factors driving accountability: institutionalized and grassroots. Institutional accountability, such as financial rewards and punishment (e.g., denial of promotion), is purposefully designed by the organization to control accountability. Grassroots accountability, by contrast, is either peer-driven or innate. Software engineers feel accountable to their peers, fostering a sense of collective responsibility. Their accountability for their outcomes, including code quality, is also driven by their intrinsic drivers, such as the desire to maintain personal standards and professional integrity. The study also identified several mechanisms to control accountability, such as performance review (institutionalized) and code review (grassroots) \citep{alami2024understanding}. We build on our earlier study's foundational understanding of accountability to examine the role of intrinsic drivers in accountability in the context of peer-led versus LLM-based code reviews. In Sect. \ref{sec:chase}, we elaborate further on the methodological and empirical relation to our previous work reported in \citep{alami2024understanding}.

In summary, related work highlights the importance of both human and social factors in shaping code review practices, from individual behaviors like cognitive load and motivation to social constructs such as trust, fairness, and conflict resolution. However, much of this work has focused on open-source settings, leaving gaps in understanding how these factors operate in institutional contexts. Furthermore, while prior studies, including our own, i.e, \citep{alami2024understanding}, have explored accountability as a broader construct in SE, how accountability manifests and influences the code review process remains underexplored. Furthermore, the specific role of intrinsic drivers in shaping accountability within peer-led and LLM-assisted reviews is yet to be investigated. In this study, we address these gaps by examining how intrinsic drivers influence accountability for code quality in peer-led and LLM-assisted reviews.

%% file: methods.tex
\section{Methods}\label{sec:methods}

\noindent We use an exploratory sequential two-phased design ($\textbf{interviews} \rightarrow \textbf{focus groups}$) \citep{creswell2017designing}. We used interviews in Phase I, which gave us rich insights into software engineers' intrinsic drivers and how they influence their sense of accountability towards the quality of their code.

In Phase II, we used user enactments with targeted focus groups in both peer-led (\textbf{RQ1}) and LLM-assisted (\textbf{RQ2}) code reviews. This design allowed us to further validate the findings from Phase I and to triangulate our findings~\citep{Storey2020} using two distinct research methods and two sources of data, ensuring empirical rigor in investigating the uncharted relationship between intrinsic drivers and accountability for code quality.

Furthermore, in Phase 1, the inductive analysis of the interview data allowed us to identify the key findings. In Phase 2, we sought to validate the earlier findings using focus groups. The secondary data analysis enhanced the earlier insights by contextualizing them within a setting close to a professional context, which the interviews alone could not fully capture.

\subsection{Phase I: Interview Study}

\begin{table*}[t!]

  \begin{center}
    \footnotesize
    \caption{Interviewees' characteristics.}
    \label{tbl:population}
    \renewcommand\arraystretch{0.80}
        
    \begin{tabular}{l|p{2.5cm}|c|c|p{4cm}|c|p{1.1cm}}
      \hline
      \textbf{\#} & \textbf{Role} & \textbf{Exp.} & \textbf{Method} & \textbf{Industry sector} & \textbf{Gender} & \textbf{Country}\\
      \hline
    
        P1 & Software Engineer & 3-5 years & DevOps & Information Technology services & Male & Germany\\
        P2 & Software Engineer & 3-5 years & Hybrid & Information Technology services & Male & UK\\
        P3 & Software developer & 9-11 years & Scrum & Robotics manufacturing & Male & USA\\
        P4 & Software developer & 9-11 years & Hybrid & Information Technology services & Male & Italy\\
        P5 & Sr. software engineer & 6-8 years & Scrum & Information Technology services & Non-binary & Germany\\
        P6 & Sr. software engineer & $>$12 years & Kanban & Banking services & Male & Canada\\
        P7 & Software developer & $<$3 years & Hybrid & Information Technology services & Male & France\\
        P8 & Sr. software engineer & $>$12 years & Scrum & Information Technology services & Female & India\\
        P9 & Sr. software engineer & 6-8 years & Scrum & Information Technology services & Male & Serbia\\
        P10 & Sr. software engineer & 9-11 years & Hybrid & Global software vendor & Male & Canada\\
        P11 & Software developer & 3-5 years & Scrum & Global software vendor & Male & UK\\
        P12 & Software engineer & $<$3 years & DevOps & Information Technology services & Female & India\\
         \rowcolor{Gray} 
        P13 & Sr. software engineer & 9-11 years & Scrum & Wearables \& IoT Technology & Female & Finland\\
        \rowcolor{Gray} 
        P14 & Software developer & 3-5 years & Hybrid & Design consultancy & Female & Canada\\
        \rowcolor{Gray} 
        P15 & Tech Lead & $>$12 years & Scrum & Technology \& engineering services & Female & USA\\
        \rowcolor{Gray} 
        P16 & Software Engineer & $<$3 years & DevOps & Information Technology services & Female & Portugal\\
    
     \bottomrule
     
    \end{tabular}
   
  \end{center}
  
\end{table*}

\paragraph*{Interviewee recruitment \& selection}\label{Interviewee_recruitment} We used Prolific\footnote{\url{https://www.prolific.co/}}, a research market platform, to recruit participants for the interviews. Recruiting qualified participants on crowd-worker platforms can be challenging \cite{alami2024you}. We followed the best practices we learned in our previous research, as explicated in \citep{alami2024understanding}. We used coding tasks and critical thinking questions grounded in the participant's experiences to evaluate their skills \citep{alami2024you}.

We sent an invitation to 562 pre-screened participants to participate in a further pre-screening survey for the interviews\footnote{All study materials, including the pre-screening survey for interviewees are available in the replication package}. Based on 170 responses we received, 16 accepted to participate. We used a broad range of criteria to select our interviewees, such as country of residence, project role, experience, gender, and accountability practices within the participant team and organization. Table \ref{tbl:population} overviews the demographics. Six respondents identified as female, one as non-binary, and nine as male.

We used purposive sampling for participant selection \citep{baltes2022sampling}. This method allowed us to use our ``judgment'' (i.e., ``the researcher can exercise expert judgment'') \citep{baltes2022sampling} in determining which participants fit our selection criteria. Even though this sampling is inherently subjective \citep{baltes2022sampling}, we mitigated potential biases in our selection by employing a diverse set of criteria \citep{baltes2022sampling}, such as participants' roles, levels of experience, industry sectors, software development methods, and geographical locations. This diversity ensures that our sample reflects a diverse range of perspectives.

For example, while P1 and P16 are both early-career software engineers working in DevOps environments, they identify with different genders and are geographically located in different countries (male from Germany and female from Portugal). Similarly, P6 and P15, both highly experienced software engineers (over 12 years of experience), differ in their roles (senior software engineer vs. Tech Lead) and industry sectors (banking services vs. technology and engineering services). In our selection process, we intentionally varied the attributes of each participant in subsequent selections to promote diversity. Each time a participant was considered for selection, we sought differences in variables such as role, experience level, methods, industry sector, gender, or geographical location from those already included in the sample. This approach allowed us to systematically build a sample that captures a diverse range of perspectives and contexts.

We also sought participants with varying degrees of accountability and practices used to control it, reinforcing the depth and breadth of our sample. For example, among our interviewees (P1, P2, and P3, as listed in Tbl. \ref{tbl:population}), all identified ``code quality'' as their primary individual accountability within their respective teams. However, they reported varying levels of accountability. P1 expressed a high level of accountability, responding with ``strongly agree'' to all accountability-related questions. In contrast, P2 and P3 displayed a moderate level of accountability, responding with ``somewhat agree'' to most of the accountability questions. Below, we highlight the recruitment process:

\begin{itemize}

    \item [-] \textbf{Initial pre-screening:} The purpose of this pre-screening phase is to evaluate potential participants' skills. Although Prolific provides an array of self-reported skills accessible in the participant's profile, they are not vetted \citep{alami2024you}. To ensure high-quality sample, researchers must prescreen their own subjects \cite{chandler2017lie}. In this pre-screening activity, we used an iterative and controlled prescreening, and task-oriented questions \citep{alami2024you}. While the iterative process (we assessed 50 responses a day) allowed us close scrutiny of the answers, we developed questions to avoid merely testing theoretical understanding, but we wanted participants to apply their skills to solve specific problems or discuss complex scenarios \citep{alami2024you}. For example, in one of the pre-screening survey, we asked: ``Please describe a specific instance where you contributed to the success of a software development project. What challenges did you face, and how did your involvement impact the project's outcome?'' This approach allowed us to ensure that selected participants had practical experience and demonstrated critical thinking skills, demonstrating genuine experience and relevant to SE. All text-free questions were scrutinized manually and ChatGPT 4.0 was used for AI-generated content. At the end of this activity, we curated a total of 562 qualified participants. This activity took place in the period of August - September 2023.

    \item [-] \textbf{Additional pre-screening:} To ensure participants with varying degrees of accountability and practices used to control it, we carried out an additional pre-screening survey. In this survey, we collected data on how accountability mechanisms shape the participant's work environment, which we did not have in previous pre-screening data. This additional data reinforced depth and breadth in our sample. For example, interviewees P1, P2, and P3, as listed in Tbl. \ref{tbl:population}, all identified ``code quality'' as their primary individual accountability in their teams. However, they reported varying levels of accountability. P1 expressed a high level of accountability, responding with ``strongly agree'' to all accountability-related questions. In contrast, P2 and P3 displayed a moderate level of accountability, responding with ``somewhat agree'' to most of the accountability questions. We received 170 responses and we selected twenty participants; twelve accepted to participate (i.e., P1-P12) (the data and findings from this sample are reported in \citep{alami2024understanding}. Due to low participation of females in the earlier selection (reported in \citep{alami2024understanding}), we invited more female to take part in the study from the pre-screened sample in the earlier pre-screening activity. We successfully recruited four additional participants (i.e., P13-P16). The first selection took place in September 2023 and the second in January 2024.

    \item [-] \textbf{Interviews:} Upon the successful selection of our participants, we interviewed P1-P12 in October 2023 and P13-P16 in January 2024.

\end{itemize}

The recruitment and data analysis processes were conducted iteratively and in parallel to facilitate the monitoring of saturation (see Sect. \ref{sec:saturation}) and to determine the appropriate sample size \citep{patton2014qualitative}. Following each interview, we carried out a preliminary analysis and performed saturation checks. These checks provided us with confidence in our sample size and the depth of evidence collected thus far.

\paragraph*{Relationship to ongoing research study}\label{sec:chase} Our Phase I interviews are part of an ongoing research project into accountability. We reported on twelve participants in \citep{alami2024understanding} and examined broadly the concept of accountability in SE. In this paper, we add four further participants to the interviews (grey rows in Table \ref{tbl:population}) and re-analyze the entire dataset, from scratch, in the context of code review, using \textbf{RQ1} and \textbf{RQ2} as new analytical lenses. We expand on the way individual and team accountability influence code quality in the context of one team-level control mechanism of accountability, code reviews. In addition, we expand on our findings by including the role that AI assistance might play in influencing accountability in the context of code review. While some intrinsic drivers (only personal standards and professional reputation) emerged as themes in the previous study \citep{alami2024understanding}, they were not a main focus of the research.

In addition, the new dataset and analysis have yielded additional themes in Phase I of the study not previously reported in \citep{alami2024understanding}, namely, professional integrity and pride in code quality. Furthermore, Phase II data has also provided new and complementary insights into the earlier findings.

Methodologically, \textit{secondary analysis}---re-analyzing existing qualitative data sets to address new research questions---is a well-accepted and sound method in qualitative research \citep{heaton2008secondary,ruggiano2019conducting}. This is permitted when the dataset adds value to the new research questions and brings deeper understanding without the need to collect new data \citep{heaton2008secondary}. We recruited four additional interviewees to the original twelve, enhancing female participation in our sample. With our new analytical lens (\textbf{RQ1}), we focused our re-analysis on code review as an accountability mechanism and how intrinsic drivers ignite a sense of accountability for code quality.

\begin{table*}[th!]
\footnotesize
    \caption{Example of Interview Questions}

    \label{tab:phase1_guide}
    \renewcommand\arraystretch{1.8}
    
    \begin{tabular}{p{2.8cm}p{10.5cm}}
    \toprule
      \textbf{Interview component} & \textbf{Examples of questions}\\
      \hline
      
      \multirow{5}{2.8cm}{\textbf{Accountability mechanisms}} &  What rules, policies, and guidelines do you have in place to ensure the outcome set for the team (e.g., quality, efficiency)?\\
      
      & Are there other team or peer expectations that aren't formal like the guidelines you mentioned?\\

      & How are these expectations controlled in your team? \\

      & How about you? How do these mechanisms influence you? Do you have an example of how you felt when you were asked to be answerable?\\ \hline
      
      \multirow{4}{2.8cm}{\textbf{Outcomes sought by accountability mechanisms}} & In your experience, how have the accountability mechanisms you mentioned influenced the desired outcomes you cited?\\

      & Have you noticed any unintended consequences or challenges associated with the implementation of these mechanisms?\\

      &  Do you have an example of yourself or a colleague who hasn't met expectations? What happened?\\ \hline

      \multirow{4}{2.8cm}{\textbf{Moderating factors}} & Are there any factors or conditions that you believe can influence the effectiveness of the accountability mechanisms you cited in your team?\\
      
      &  How do team dynamics and individual characteristics (e.g., experience, role) influence the impact of accountability mechanisms?\\

      &  Can you share an example of how you felt accountable and why?\\
      
     \bottomrule
     
    \end{tabular}
  
\end{table*}

\paragraph*{Data collection} We used semi-structured interviews to collect Phase I data. We remained flexible during the interviews, aligning the predefined interview questions with the flow of the discussion.

Our interviews covered background, a general understanding of accountability in their teams, and then questions to identify how accountability mechanisms impact the desired outcomes. In line with \textbf{RQ1} objectives, we asked questions to understand how personal traits interplay with the feeling of accountability towards a team's and individual's outcomes. In line with best practices in qualitative research, we did not ask directly about intrinsic drivers to avoid desirability bias and self-perception inaccuracies \citep{stone1999science,podsakoff2003common}. Instead, we sought to identify possible moderating factors of accountability, such as experience, role hierarchy, and personal traits. Before concluding the interview, we asked how the interviewee's team implements effective accountability mechanisms. Table \ref{tab:phase1_guide} documents examples from the interview guide. A detailed and complete guide is available in the replication package (Sect. \ref{sec:replication}).

We anchored the design of the interview guide in the existing theoretical foundation drawn from work in social sciences (see Sect. \ref{sec:theory}). For instance, the theory of accountability proposes the use of either formal or informal methods to control it. To align with this perspective, we designed questions to identify formal and informal accountability practices, such as ``What rules, policies, and guidelines do you have in place to ensure the outcome of discussed (e.g., quality, efficiency)?'' and ``Are there any other expectations set by the team or your peers, but they are not necessarily formal like the guidelines you mentioned?''

We structured the guide into five sections: after the introduction to the interview, section one sought to capture data on the background of the interviewee and a general understanding of how accountability manifests in their teams. Section Two sets out to identify accountability mechanisms and the outcomes they seek to control in the interviewee's team. Section Three delves into how the discussed accountability mechanisms impact the desired outcomes. Section Four aimed to reveal any moderating factors or conditions (e.g., roles or interpersonal relations at work) that may influence the effectiveness of accountability mechanisms. Before concluding the interview, Section Five identified best practices used by the interviewee's team to implement effective accountability mechanisms.

Our interviewees were distributed geographically. We conducted the interviews using Zoom to accommodate the distributed nature of our sample. Even though ten of our participants are non-native English speakers, we did not encounter any language challenges. This is likely because our interviewees are highly educated and from countries where English is widely spoken as a second language, e.g., India, Finland, and Germany. The interviews lasted 40-60 minutes with a total of 13h40m of audio. The audio generated a total of 189 pages verbatim after transcribing. The first author conducted the interviews in October 2023 (for P1-P12) and in January 2024 (for P13-P16). We used Otter.ai,\footnote{\url{https://otter.ai/}} an online transcription tool, to transcribe the audios. We paid \textsterling30 to each interviewee for their participation.

\paragraph*{Saturation} \label{sec:saturation} To ensure that the sample size is adequate, we monitored data saturation \citep{morse2004theoretical,aldiabat2018data} throughout our iterative analysis process. During this iterative process, we continuously compared data and emerging themes, switching back and forth between data collection and analysis \citep{bowen2008naturalistic}. This process allowed us to observe when some of our themes reoccur strongly in the data. This iterative approach also allowed us to monitor and assess the sample size, i.e., determine whether more participants are required. We were able to conclude data collection with 16 participants, and at the same time, no new themes emerged from the analysis. Personal standards and reputations reached saturation at 12 interviews, pride and professional integrity at 16. The additional four interviews collected for this study accentuated intrinsic drivers and provided greater depth to the themes.

In this process, we combined iterative analysis and saturation monitoring, which allowed us to determine that 16 participants provided sufficient data to address our RQs comprehensively, as no new themes emerged beyond this point \citep{guest2006many}. Our approach aligns with best practices in qualitative research, we determined our sample size by the depth and richness of the data we collected than a pre-determined numerical thresholds \citep{guest2006many}.

As this study represents a re-analysis with new research questions, we recommenced saturation monitoring from the first interview to align with the updated analytical lens (this study's \textbf{RQ1} and \textbf{RQ2}) \citep{heaton2008secondary}. Saturation monitoring followed an iterative process, documented using a tracking spreadsheet (available in the replication package, see Sect. \ref{sec:replication}).

\paragraph*{Member checking} Upon the completion of the analysis, we conducted member checking~\citep{miles2014qualitative,birt2016member}. We sought feedback from our interviewees on our findings. We documented our themes in a separate document for each interviewee. Then, we shared the link with each interviewee in Prolific and asked them to comment and provide feedback. Twelve interviewees provided feedback, and four opted not to provide feedback. No major objections to our interpretations were reported. Some interviewees requested further clarifications prior to providing their support for the findings. We paid an additional \textsterling5 for this effort.

\subsection{Phase II: Focus groups}\label{sec:phase_2}

\afterpage{

\begin{landscape}

\renewcommand\arraystretch{1.8}

\footnotesize

    \begin{longtable}{lp{3.5cm}p{6.5cm}p{5.5cm}p{2cm}}
    \caption{Configuration of the Four Focus Groups}
     \label{tbl:focus_groups}\\
     \toprule

     \textbf{\#} & \textbf{Scenario} & \textbf{Research Objective} & \textbf{Participants} & \textbf{Main Variable} \\
        \midrule

        \multirow{6}{0.5cm}{\textbf{FG1}} & \multirow{6}{3cm}{Role hierarchy} & To explore how hierarchical dynamics (e.g., junior developers reviewing senior developers' code) influence the persistence of intrinsic drivers like pride in code quality and professional integrity. This setup assesses whether intrinsic motivations are moderated by authority bias and social influence in hierarchical settings, as participants navigate power dynamics during reviews. This scenario helped validate whether intrinsic drivers such a personal standards are upheld in the face of authority. & Two junior and one mid-career developer as reviewers, and three code authors (two senior developers, and a team lead) & Role hierarchy.\\
        
        \hline
        
        \multirow{6}{0.5cm}{\textbf{FG2}} & \multirow{6}{3cm}{Public reviews} & Impact of publicly available review comments on accountability. The setup used public visibility in simulated review platforms to explore whether participants adapted their feedback style and accountability practices. This scenario aligns with prior findings that public feedback increases performance due to visibility but may compromise candidness and individual accountability due to fear of judgment and retaliation. By observing participants' behaviors and reflections on providing feedback in this context, the study captured whether their intrinsic drivers, such as professional reputation, were heightened or compromised in public review settings. & 3 developers (mixed roles seniority) as reviewers, 3 developers (mixed roles seniority) as code authors & Public review. \\
        
        \hline
        
        \multirow{6}{0.5cm}{\textbf{FG3}} & \multirow{6}{3cm}{Cross-team review} & The focus group design introduced complex interdependencies between participants assigned to different ``teams'' contributing to the same codebase. This cross-team reviews was informed by complexity theory, which suggests that accountability becomes diffused in such settings. Observing how participants balanced individual accountability with the need for collective ownership helped validate whether intrinsic drivers persisted or were overridden by external pressures in collaborative settings. & Mixed team roles (3 senior developers, and 2 mid-career developers) as reviewers and authors & Cross-team reviews. \\
        
        \hline
        
        \multirow{6}{0.5cm}{\textbf{FG4}} & \multirow{6}{3cm}{Complex code module review with varying quality levels} & To investigate the persistence of intrinsic drivers like personal standards and professional integrity in reviewing technically complex and variable-quality code. This scenario tests how participants' accountability adapts to cognitively demanding and high-pressure review conditions. For example, when faced with complex code and high cognitive effort, both individual and collective accountability may reduce under high cognitive demand. & Three developers (mixed roles) as reviewers, three developers (mixed roles) as code authors & Code complexity. \\
        
        \bottomrule
    
    \end{longtable}

\end{landscape}

}

In Phase II we conducted four focus groups with five to six participants in each group (see Tbl. \ref{tbl:focus_groups_population}). This method complements and elaborates Phase I findings in a more interactive and collaborative environment. 

Self-reported data is subject to potential biases, such as social desirability bias, recall bias, and self-perception inaccuracies, resulting in over- or under-reporting of behaviors and attitudes \citep{podsakoff2003common}. For example, participants might unintentionally or intentionally report data and events in a more favorable light, either to themselves or in a way that aligns with what they perceive the researcher favors \citep{stone1999science,podsakoff2003common}. Such behavior may impact the reliability of the data. In addition, self-reported data often fail to capture the complexity and dynamics of group interactions and decision-making processes, which we have observed in this study. For example, while in Phase I we learned that individual accountability is the main driver for pursuing higher code quality, Phase II focus groups revealed that the collaborative aspects and peers' influence during code review shift the sense of accountability from an individual-level to a collective-level. This intricate detail brought a more comprehensive understanding of how accountability may take different levels depending on the activity within the social and collaborative contexts of code review.

By incorporating focus groups in Phase II, we managed to capture the social and collaborative aspects of code review processes and observe participants' behaviors and interactions in real-time, providing richer and more nuanced data. This methodological triangulation \citep{perlesz2003methodological} enhances the validity and reliability of our findings. 

However, we acknowledge that focus groups, as a method, also carry inherent biases, such as social pressure and groupthink, which may influence participants' responses and participation \citep{o2018use}. We employed several techniques to mitigate these biases. The moderated focus groups ensured inclusive contributions from all participants. The discussions used open-ended questions to prompt participants to share their viewpoints, reducing the likelihood of convergence toward a single opinion. At the end of each focus group, all participants were encouraged to provide individual reflections to capture perspectives uninfluenced by group dynamics. Additionally, we employed participant-authored code to lessen the artificiality often associated with focus group settings. This approach facilitated the observation of genuine interactions and decision-making processes, as participants engaged more naturally with their own code. Best practices in focus group and simulation studies highlight the importance of using contextually relevant and participant-generated materials to enhance ecological validity and minimize biases inherent in simulated settings \citep{gaba2004future}.

We organized our two-hour focus groups into two parts: the first hour was dedicated to peer-led reviews (\textbf{RQ1}) and the second hour to LLM-led reviews (\textbf{RQ2}).

\subsubsection{Peer-led reviews}

We designed the focus groups to enable the ``enactment'' of our participants. In user enactments, researchers design a physical and social context to simulate a future situation \citep{odom2012fieldwork}. Users are asked to enact loose scenarios of situations familiar to them \citep{odom2012fieldwork}. This design allows the researcher to observe and probe participants, grounding the discussion in a context similar to the participants' professional experiences \citep{odom2012fieldwork} and enhancing the authenticity of qualitative data collected \citep{carroll2003making}.

\paragraph*{Configuration} We designed four different configurations for our focus groups' scenarios, as illustrated in Tbl. \ref{tbl:focus_groups}. Table \ref{tbl:variables_description} documents the variables we used to configure each focus group. By incorporating a diverse range of variables, such as role hierarchy, review visibility, urgency, code complexity, and team collaboration, we aimed to test software engineers' intrinsic drivers influencing their sense of accountability for code quality in settings that resonate with their real-life circumstances. For example, while the first scenario focuses on evaluating the \textit{resilience} of reported intrinsic drivers like pride in code quality and professional integrity when junior developers review their senior counterparts' code, the third scenario does the same in a \textit{cross-team review} setting.

We designed the first focus group, \textbf{FG1}, to test the persistence of intrinsic drivers, such as maintaining personal standards for code quality, in a hierarchical setting. Several studies in SE reported the impact of role seniority on team's dynamics \citep{cunha2021code,sadowski2018modern,bacchelli2013expectations}. In summary, these studies report that senior developers influence the code review process positively by providing mentorship, maintaining high standards, and enriching the learning experience of juniors \citep{cunha2021code,sadowski2018modern,bacchelli2013expectations}. However, authority bias theory suggests that employees defer decisions to those perceived in higher positions or more knowledgeable, reducing their sense of individual accountability \citep{kipnis1972does,milgram1963behavioral}. Similarly, social influence theory posits that individuals are more likely to conform to the opinions and behaviors of those they perceive as more knowledgeable or authoritative \citep{cialdini2004social,asch2016effects}. In a code review context, junior developers might align their actions and decisions with those of senior developers, thereby reducing their sense of individual accountability for the code they are asked to review or have authored.

The aim of \textbf{FG2} is to examine the influence of how public feedback settings impact the sense of accountability for code quality among team members. Impression management theory posits that in a public setting, individuals may attempt to control their behaviors to influence impressions others form of them \citep{bolino2016impression}. Such behavior can lead to appearing in conformance to social norms \citep{bolino2016impression}. In the context of code review, public feedback may incentivize individuals to perform well due to broader visibility, which may heighten accountability. Fear of negative evaluation can also influence behaviors when feedback is shared publicly \citep{watson1969measurement}. Developers may avoid conveying negative comments to peers to mitigate the risk of damaging relationships or future retaliations \citep{kocovski2000social}, which may result in compromised levels of accountability for code quality. These theoretical perspectives have been echoed in SE literature. For example, Rigby et al. report that public peer reviews enhance performance due to the social facilitation effects \citep{rigby2013convergent}. Bosu et al. suggest that developers align their comments with the group's norms and expectations \citep{bosu2015characteristics}, indicative of impression management behaviors in code review.

\begin{table}[t!]

    \caption{Variables Used to Configure Focus Group Scenarios}
    \label{tbl:variables_description}
    \centering
    
    \begin{tabular}{rp{10.5cm}}
    
        \toprule
        \textbf{Variable} & \textbf{Description and Rationale} \\
        
        \midrule
        Role hierarchy & Refers to the structured levels of authority, seniority, and responsibility within the focus group participants.\\

        Public reviews & Code reviews conducted in a open tool with broader visibility, e.g., review comments made available on GitHub or Bitbucket for for other than the group members to consult. \\

        Cross-team reviews & This variable reflects the need to review cross teams when a codebase involves multiple teams collaboration. \\

        Code complexity & Presents the technical challenge presented by the code under review. We used this variable to to present code with varying level of complexity in our scenarios. \\
        
        \bottomrule
        
    \vspace{-0.5cm}
    
    \end{tabular}

\end{table}

\textbf{FG3} reflects the need to review cross teams when a codebase involves multiple teams collaboration. In a cross-team reviews, dealing with complex codebases can often involve critical interdependencies, thereby heightening accountability. Complexity theory suggests that complex system often exhibit nonlinearity, meaning that the effect is not always proportionate to a cause \citep{byrne2002complexity,larsen2013complexity}. In a nonlinear system, a trivial change in one variable can cause significant implications downstream \citep{byrne2002complexity,larsen2013complexity}. This uncertainty may require a clear and higher accountability to ensure that all aspects of the code are thoroughly reviewed before integration. In complex systems, accountability is also difficult to
pinpoint \citep{kacianka2021designing}. For example, a ``pilot error'' is not simply a failure of the pilot, but rather a complex interplay of the humans and the technical systems \citep{kacianka2021designing}. To test the manifestation and the persistence of Phase I findings of accountability for code quality in a complex setting, we assigned \textbf{FG3}' participants to three different groups with defined interdependencies in the code subject to review (see our focus group scenario designs in the replication package, Sect. \ref{sec:replication}, for further details).

While \textbf{FG3} is to test accountability when the system is socially complex, multiple teams managing the same codebase with interdependencie, \textbf{FG4} aims to explore accountability in technically complex scenarios within the same team. Transactive memory systems (TMS) theory suggests that in such environments, the division of cognitive labor and shared knowledge among team members are crucial for effective coordination and accountability \citep{wegner1987transactive}. To evaluate Phase I findings regarding accountability in the face of technical complexity, we presented \textbf{FG4} participants with complex code snippets of varying quality. This allowed us to test how well our participants maintain accountability under more technically challenging conditions.

To mimic real-world conditions and capture discussions in relatable SE settings, we designed each focus group scenario to reflect typical challenges faced by software engineers in professional settings. For instance, in \textbf{FG1}, the hierarchical setup of juniors reviewing seniors' code was structured to replicate dynamics where authority and experience levels intersect, creating opportunities to observe potential authority bias. \textbf{FG2} aims to test whether accountability intensifies when reviews are visible to broader audiences outside the team. Similarly, in \textbf{FG3}, the cross-team review configuration was designed to simulate the interdependencies often seen in large-scale collaborative projects, where multiple teams contribute to a shared codebase. \textbf{FG4} intended to also evaluate whether accountability shifts when complexity is higher. Existing literature also informed these scenarios.

Admittedly, our scenarios do not represent every possible SE condition, nor does their design capture the true complexity of real-life circumstances. However, we selected and designed these scenarios based on potential changes in the social setting introduced by the variables (e.g., hierarchy, and public reviews), which may compromise the individual and collective levels of accountability. Our choices are also grounded in theoretical perspectives. We further discuss this threat to validity in Sect. \ref{sec:trust}.

\paragraph*{Code generation} We opted for Python to write code snippets subject to reviews in the focus groups. The choice of Python is rationalized by its widespread acceptance and use in the software development community. Python's readability and straightforward \citep{van1995python} syntax make it an ideal choice to illustrate potential quality issues in a review setting. This choice also facilitates better focus on quality issues rather than the intricacies of the programming language \citep{dagenais2011recommending}, which may distract the focus of the discussions. Dagenais and Robillard, and Baxter and Sommerville studies suggest that using familiar programming languages and frameworks facilitates better understanding and focus in the scope of software engineering studies \citep{dagenais2011recommending,baxter2011socio}.

For the code snippets, we aimed for realistic scenarios. Scenario-based design literature recommends maintaining realistic scenarios in qualitative research \citep{carroll2003making}. This approach does not only mitigate the risk of drifting away from the actual scope of the study but also ensures that the data collected facilitates a shared understanding with the participants \citep{carroll2003making}. This approach was echoed in nursing education literature \citep{jeffries2005framework,shin2015effectiveness,cant2010simulation}. Realistic simulation scenarios are used to help students bridge the gap between theoretical knowledge and practical skills \citep{jeffries2005framework,shin2015effectiveness,cant2010simulation}. Realistic scenarios also improve the ecological validity of the design and its setting, meaning that the data collected is more relatable \citep{gaba2004future}. Table \ref{tbl:code_snippets} summarizes the code snippets for all our focus groups; further details are available in the replication package (see Sect. \ref{sec:replication}). The ''Ref.'' column lists the snippet file names as they appear in our replication package. All snippets were written by the first author and reviewed by the third author prior to the focus groups taking place.

\afterpage{

\begin{landscape}

\renewcommand\arraystretch{1.0}

\footnotesize

    \begin{longtable}{lp{3cm}p{7.5cm}p{6.2cm}}
    \caption{Python Code Snippets Used in Focus Groups}
    \label{tbl:code_snippets}\\
     \toprule

        \textbf{Focus Group} & \textbf{Ref.} & \textbf{Python Code Snippets} & \textbf{Main Quality Issues} \\
        
        \midrule

         \multirow{4}{2.5cm}{\textbf{FG1}- Role hierarchy} & FG1\_Scenario\_1\_Snippet\_1 & \multirow{2}{7.5cm}{This code snippet processes a list of data, doubling the value of integers.} & 1. Inefficient handling of different data types.\\

         &  &  & 2. Lack of type hinting and docstrings.\\

        &  &  & 3. No exception handling for unexpected data types.\\
        
        \cline{3-4}
        
         & FG1\_Scenario\_1\_Snippet\_2 & \multirow{2}{7.5cm}{This code snippet retrieves user data from an API using the requests library.} & 1. Limited error handling; only catches RequestException.\\

         &  &  & 2. Hard coded URL without parameter validation.\\
        
        \hline
        
        \multirow{8}{2.5cm}{\textbf{FG2}- Public reviews} & \multirow{2}{2.5cm}{FG2\_Scenario\_2\_Snippet\_1} & \multirow{2}{8.5cm}{This code snippet extracts the file extension from a filename.} & 1. Does not handle filenames without extensions.\\

         &  &  & 2. No validation for input types.\\
        
        \cline{3-4}
        
         & \multirow{4}{2.5cm}{FG2\_Scenario\_2\_Snippet\_2} & \multirow{4}{8.5cm}{Authenticates a user against a user database.} & 1. Passwords are stored and checked in plain text, which is insecure.\\

         &  &  & 2. Should use hashed passwords and secure comparison methods.\\

         &  &  & 3. No exception handling for dictionary key errors.\\

         \cline{3-4}

        & \multirow{3}{2.5cm}{FG2\_Scenario\_2\_Snippet\_3} & \multirow{3}{8.5cm}{Aggregates data records by category.} & 1. Potential for key errors if category or data fields are missing in any record.\\

         &  &  & 2. Could validate input data before processing.\\

         &  &  & 3. Inefficient handling of large datasets.\\
        
        \hline
        
        \multirow{9}{2.5cm}{\textbf{FG3}- Cross-team reviews} & \multirow{2}{2.5cm}{FG3\_Scenario\_3\_Snippet\_1} & \multirow{2}{8.5cm}{Executes a database query and returns the results.} & 1. No error handling for connection establishment failures.\\

         &  &  & 2. Could have handled specific database errors more robustly.\\
        
        \cline{3-4}
        
         &  \multirow{2}{2.5cm}{FG3\_Scenario\_3\_Snippet\_2} & \multirow{2}{8.5cm}{Asynchronously processes data from a queue.} & 1. Exception handling is minimal; could be improved to handle specific errors.\\

         &  &  & 2. Should ensure that data queue operations are safe and handle timeouts.\\

         \cline{3-4}

        &  \multirow{2}{2.5cm}{FG3\_Scenario\_3\_Snippet\_3} & \multirow{2}{8.5cm}{Predicts an outcome using a machine learning model.} & 1. Retraining the model within the prediction function is inefficient.\\

         &  &  & 2. Should handle model errors separately and more efficiently.\\
        
        \hline

        \multirow{4}{2.5cm}{\textbf{FG4}- Complex code} & \multirow{2}{3cm}{FG4\_Scenario\_4\_Snippet\_1} & \multirow{2}{8.5cm}{This snippet processes data from a file asynchronously.} & 1. No validation for file content structure.\\

        &  &  & 2. Lack of proper error handling for file operations.\\
        
        \cline{3-4}
        
        &  \multirow{2}{3cm}{FG4\_Scenario\_4\_Snippet\_2} & \multirow{2}{8.5cm}{This snippet extracts links from a webpage using BeautifulSoup.} & 1. No validation for input URL.\\

        &  &  & 2. Lack of error handling for HTTP and parsing errors.\\

        \bottomrule
    
    \end{longtable}

\end{landscape}

}

\subsubsection{LLM-assisted reviews} We asked the participants in the focus groups to volunteer to submit Python source code that they had authored. We asked for professionally authored code or accepted and merged contributions to open-source projects. We used ChatGPT 4 for the LLM reviews. We used this prompt to generate a ChatGPT review: ``You are a Python software developer expert. Conduct a code review of the attached code and provide thorough feedback to the author of the code.'' Eighteen participants submitted their Python code, and five opted not to; still, they contributed to the discussion part of the LLM review. The five participants chose not to submit code, citing concerns of making it publicly available, given it was authored in proprietary contexts.

\subsubsection{Participant recruitment and selection} We used UpWork\footnote{\url{https://www.upwork.com/}}, a marketplace for freelancing work. Although we used Prolific successfully in Phase I, the effort of pre-screening and qualifying candidates was lengthy. In UpWork, by comparison, potential participants disclose their GitHub and LinkedIn profiles as part of the recruitment process to qualify for the study, which has allowed for fast and accurate vetting. 

We posted a job description for the study (available in the replication package). We asked prospects to submit their GitHub and LinkedIn profile links or a copy of their resumes to help us qualify their skills. We received 56 submissions. Based on the requirements we set in the design of our focus groups, we selected 24 participants, but only 23 participated. We used similar requirements as per Phase I. We aimed for a diverse sample while aligning with our focus group design. For instance, for FG1, we sought participants with varied seniority levels as per the scenario design. We also examined potential participants' profiles to ensure recent hands-on experience in code review and proficiency in Python as evidenced by GitHub contributions or their resumes.

Table \ref{tbl:focus_groups_population} documents the characteristics of the focus group participants, the corresponding focus group they took part in, and the roles they assumed in the scenario (e.g., junior developer, team lead, etc.). For example, P17 assumed the role of a junior developer in the first focus group (see Tbl. \ref{tbl:focus_groups}). We were only able to recruit one female participant (P26). We paid each participant \$60 for their participation in the 2-hour focus group.

We also aimed to match the participants' career and experience levels with their roles in the focus groups. For example, P17 played the role of a junior developer in FG1 and also worked as a junior software engineer at the time of recruitment. Participants are more likely to act naturally within familiar professional contexts when asked to engage in a setting that closely mirrors their professional collaborative settings \citep{jeffries2005framework,gaba2004future}.

\begin{table*}[th!]
\footnotesize
    \caption{Example of Focus Group Interview Questions}

    \label{tab:phase2_guide}
    \renewcommand\arraystretch{1.8}
    
    \begin{tabular}{p{2.8cm}p{10.5cm}}
    \toprule
      \textbf{Discussion component} & \textbf{Examples of questions}\\
      \hline
      
      \multirow{4}{2.8cm}{\textbf{Peer-led review (FG1)}} &  What thoughts or considerations were in your mind while reviewing the code?\\

      & Can you share your thoughts or feelings during the junior developers' review?\\

      & Were there moments in the review that stood out to you? Why?\\

      & How do you typically react to feedback in a similar hierarchical team setting?\\ \hline
      
      \multirow{4}{2.8cm}{\textbf{LLM-assisted review}} & What was your initial reaction to the LLM's feedback on your code?\\

      & How does the idea of having an LLM review your code impact your approach to maintaining high personal standards?\\

      & Do you believe that LLM-based reviews can accurately reflect and potentially influence your professional reputation among peers? Why or why not?\\ 
      
     \bottomrule
     
    \end{tabular}
  
\end{table*}

\afterpage{

\begin{landscape}

\renewcommand\arraystretch{1.5}

\footnotesize

    \begin{longtable}{clp{2.5cm}lllp{3.1cm}l}
    \caption{Participants Characteristics and Their Distribution Across Focus Groups}
     \label{tbl:focus_groups_population}\\
     \toprule

         \textbf{Group} & \textbf{ID} & \textbf{Role} & \textbf{Experience} & \textbf{Industry} & \textbf{Country} & \textbf{Role in the Focus Groups} & \textbf{Review/Author}\\
         \hline

         \multirow{6}{*}{FG1} & P17 & Jr. Software Engineer & 3-5 years & Mobile Apps development & Hungary & Junior Developer \#1 & Reviewer\\ 
                              & P18 & Lead Java Developer & $>$12 years & Information Technology services & USA & Team Lead & Author\\ 
                              & P19 & Sr. Software Engineer & 9-11 years & Financial services & Bulgaria & Senior Developer \#1 & Author\\ 
                              & P20 & Software Engineer & 5-9 years & Technology startup & Germany & Developer & Reviewer\\ 
                              & P21 & Jr. Software Engineer & 3-5 years & Global software vendor & UK & Junior Developer \#2 & Reviewer\\ 
                              & P22 & Tech Lead & $>$12 years & Motor vehicle manufacturing & Poland & Senior Developer \#2 & Author\\  \hline
         \multirow{6}{*}{FG2} & P23 & Software Engineer & 3-5 years & Information Technology services & Dominican Republic & Junior Developer \#1 & Author\\ 
                              & P24 & Software Engineer & 3-5 years & Information Technology services & India & Junior Developer \#2 & Reviewer\\ 
                              & P25 (Feedback)& Sr. Software Engineer & $>$12 years & IT Services and Consulting & Canada & Senior Developer \#1 & Reviewer\\ 
                              & P26 (Feedback)& Sr. Software Engineer & 9-11 years & Information Technology services & India & Senior Developer \#2 & Author\\ 
                              & P27 (Feedback)& Sr. Software Engineer & $>$12 years & Technology startup & US & Senior Developer \#3 & Author\\ 
                              & P28 (Feedback)& Tech Lead & $>$12 years & Broadcast  \& media production & US & Senior Developer \#4 & Reviewer\\  \hline
         \multirow{6}{*}{FG3} & P29 & Software Engineer & 9-11 years & Medical equipment manufacturing & Hungary & Developer from Team A & Reviewer\\ 
                              & P30 (Feedback)& Software Engineer & 9-11 years & eCommerce software development & UK & Developer from Team B & Author\\ 
                              & P31 & Tech Lead & $>$12 years & Higher education & Lithuania & Tech Lead from Team A & Author\\ 
                              & P32 & Sr. Software Engineer & $>$12 years & IT Services and Consulting & Serbia & Sr. Developer from Team A & Author\\ 
                              & P33 & Sr. Software Engineer & $>$12 years & Software development services & India & Sr. Developer from Team C & Reviewer\\ \hline
         \multirow{6}{*}{FG4} & P34 & Sr. Software Engineer & $>$12 years & Public policy advisory & UK & Senior Developer \#1 & Author\\ 
                              & P35 & Software Engineer & 9-11 years & Global software vendor & UK & Developer \#1 & Reviewer\\ 
                              & P36 (Feedback)& Software Engineer & 6-8 years & IT Services and Consulting & France & Developer \#2 & Reviewer\\ 
                              & P37 & Sr. Software Engineer & $>$12 years & Software development service & Brazil & Senior Developer \#2 & Author\\ 
                              & P38 & Sr. Software Engineer & $>$12 years & Information Technology services & Serbia & Senior Developer \#3 & Author\\ 
                              & P39 & Sr. Software Engineer & $>$12 years & Software development service & Canada & Senior Developer \#4 & Reviewer\\ 
                              \bottomrule

    \end{longtable}

\end{landscape}

}

\subsubsection{Data collection} We designed our scenarios for groups of six participants. Prior to each focus group taking place, we sent detailed instructions to each participant describing the focus group process and the code they had to review (the instructions sent to the participants are shared in the replication package). 

To ensure that the code under review would be familiar and accessible to all participants, we asked them to answer a pre-screening question related to their competence in programming with Python (see UpWork job ad in the replication package).

All focus groups were conducted online using Zoom. One participant did not show up to \textbf{FG3}, yet we managed to conduct that group with minimal impact on its purpose. All focus groups were audio recorded and transcribed using Otter.ai. The first author facilitated the focus groups in February 2024. The focus group's audios generated between 24-27 pages of verbatim each after transcription, presenting approximately a total of seven hours of audio recordings.

We deemed four groups with varying scenarios sufficient to cover a spectrum of settings. We prepared a discussion guide for each focus group to structure the dialogue; however, we remained flexible during the facilitation and prompted the participants with follow-up questions when necessary. Table \ref{tab:phase2_guide} documents an example of questions used to guide the discussion. A detailed and complete guide is available in the replication package (Sect. \ref{sec:replication}).

We used a pre-focus group questionnaire to collect data on participants' intrinsic drivers, i.e., pride, integrity, reputation, and upholding personal standards, and how they influence their accountability for code quality. This methodological choice was strategically implemented to mitigate the risk of social desirability bias \citep{furnham1986response} and self-censorship \citep{yanos2008false} within the focus group discussions. We used the collected data as a reference point to prompt participants during the focus groups, encouraging them to share their true and authentic thoughts. For example, while most participants in the pre-focus group questionnaire reported pride in code quality as a key driver for their accountability to meet quality expectations, the focus group discussion hinted that their pride becomes less relevant in a group setting. When the researcher prompted the participant to explain the misalignment between what they previously reported, we learned that pride is tuned down to foster group consensus. This method is aligned with best practices in qualitative research, which suggest using pre-collected data and previously expressed views to reduce social desirability bias by encouraging honest and reflective responses \citep{king2000social,tourangeau2007sensitive}.

When the participants' discussions diverged from the pre-reported answers, we asked them to provide an explanation. For instance, when some participants' inputs during the focus group discussion did not align, the moderator prompted them to clarify. This was more prominent for ``pride in code quality'' and ``professional integrity.'' Then, we learned from the participants' explanation that their intrinsic drivers are regulated to accommodate the collective accountability and consensus.

\paragraph*{Peer-led review} We assigned the roles to the participants and asked them to prepare their reviews prior to the focus groups. The first hour of the session was used to share and discuss the feedback. Then, the researcher prompted the participants to elaborate and discuss how they felt accountable for the quality of the code. We used a pre-defined discussion guide across all sessions, which is available in the replication package. However, we allowed for some fluidity in the discussion to facilitate a natural conversation, where participants could express their thoughts freely. This approach is aligned with qualitative research best practices, which recommend a balance between structured guidance and flexible exploration \citep{patton2014qualitative}.

\paragraph*{LLM review} In the second part of the focus groups, we shared the reviews generated by ChatGPT with the participants, then asked them to reflect on the comments. Similarly to the first part, the researcher guided the discussion using a pre-defined guide for this part of the session.

\subsubsection{Feedback session} After the analysis of Phase II data was completed, we conducted a feedback session on our findings~\citep{alami2022scrum} with six participants from Phase II, annotated with asterisks in Tbl. \ref{tbl:focus_groups_population}. The purpose of this session is to collect feedback on the findings and ensure they resonate with our participants \citep{birt2016member}. During the session, we presented the findings in a scenario-like manner to ease comprehension and relatability to the participants. Our findings received support from the participants, and we did not deem revising our analysis necessary. The transcript of the session and the scenarios used are available in the replication package. The session was carried out by the first author on the first week of March 2024.

\subsection{Data Analysis and Integration} 

\begin{table*}[th!]
\footnotesize
    \caption{Example of \textbf{RQ1} Pattern Codes}

    \label{tab:themes}
    \renewcommand\arraystretch{1.8}
    
    \begin{tabular}{p{3.2cm}p{2.8cm}p{6.5cm}}
    \toprule
      \textbf{Pattern codes} & \textbf{First Cycle codes} & \textbf{Examples from the data}\\
      \hline
      
      \multirow{8}{*}{\textbf{Code quality}} & Code quality  & \emph{``So I have to write good quality code, good code, the good code, which is working and my teammates are happy with''} (P11).\\
      
      & Code maintainability & \emph{``So talking about that if I am accountable to the code quality, because suppose some other developer takes over me, or maybe someone has to extend the functionality, it should be very quick. And like it should be in a modular fashion. So, that is why we focus on code quality ...''} (P12). \\
      
      & Code readability  & \emph{``... sometimes if my code is reviewed by my senior engineer, sometimes even if the rule is not followed, but if code is readable. And if there is no very big, silly mistakes, you can say they pass the code reviews.''} (P11). \\
      \hline
      
      \multirow{2}{4.1cm}{\textbf{Individual accountability}} & Accountable for code  & \emph{``... you have to be accountable for your code ...'' P(6).} \\
      
      & Accountability to self  & \emph{``You know, just accountability towards myself''} P(7).  \\ \hline

      \multirow{6}{*}{\textbf{Intrinsic drivers}} &  Professional integrity & \emph{`` ... Integrity as in a general sense ... I think [it's s] core value for me not doing something the wrong way''} (P7). \\
      
      & Professional reputation & \emph{``If I'm working somewhere, I think that I should have a good image that is a good employee is getting good results is writing good quality code'' (P11).}  \\
      
      & Pride in code quality & \emph{``I believe the primary motivation for achieving high-quality code comes from within. It's about the personal satisfaction and pride I feel in my work'' (P14).}  \\
      
     \bottomrule
     
    \end{tabular}
  
\end{table*}

We used the same process to analyze both interview and focus group datasets, first analyzing interviews, then focus groups as per our two-phased design. Our analysis employed \textbf{inductive thematic analysis}, following the guidelines from Miles et al.~\citep{miles2014qualitative} and Salda{\~n}a~\citep{saldana2021coding} to analyze the interview data. The guidelines recommend two phases: (1) \textit{First Cycle} and (2) \textit{Second Cycle} \citep{miles2014qualitative,saldana2021coding}. The iterative approach of these guidelines allowed us to move between the data and emerging codes, which facilitated better monitoring of saturation.

\paragraph*{First Cycle} In this phase of the coding, data ``chunks'' or segments are assigned labels or codes that represent their meanings. We used an inductive coding approach to derive codes directly from the data without imposing any preconceived notions. This approach allowed us to gain data-driven insights contextualized to SE.

The first author led the first phase of coding and induced a preliminary set of codes. The initial list of codes was then reviewed by the second and third authors, who provided feedback, proposed new codes, and suggested modifications to existing labels. Following this review, the first author integrated the feedback and proposed a final list of codes. This collaborative and iterative process allowed us to refine our initial coding efforts and also ensured a robust and consistent coding scheme. This enhances the credibility and analytical process of our conclusions \citep{miles2014qualitative}.

\paragraph*{Second Cycle} In this second phase of coding, we evolved the detailed list of the First Cycle codes into a consolidated thematic structure \citep{saldana2021coding}, known as Pattern Codes~\citep{miles2014qualitative}. This condensing exercise is based on codes that share themes, patterns, or logical characteristics. The first author led this phase, with subsequent reviews and input from the second and third authors to ensure a unified perspective and reach consensus. Table \ref{tab:themes} documents examples of some Pattern Codes and their corresponding \textit{First Cycle} code and quotes from the data.

To mitigate the risk of social desirability bias (responses favored by the researchers and other participants) \citep{furnham1986response}, we explicitly did not use Phase I Pattern Codes in the design of the discussion guides for Phase II (focus groups). During focus group discussions, without being led towards specific responses, participants were encouraged to reflect on how their personal standards, pride, reputation, and professional integrity affect their approach to code review and feedback, based on their responses in a pre-focus group questionnaire (see Sect. \ref{sec:phase_2}).

\begin{figure*}[!t]
    \includegraphics*[trim=1cm 1.5cm 1cm 1cm, clip, width=1.0\textwidth]{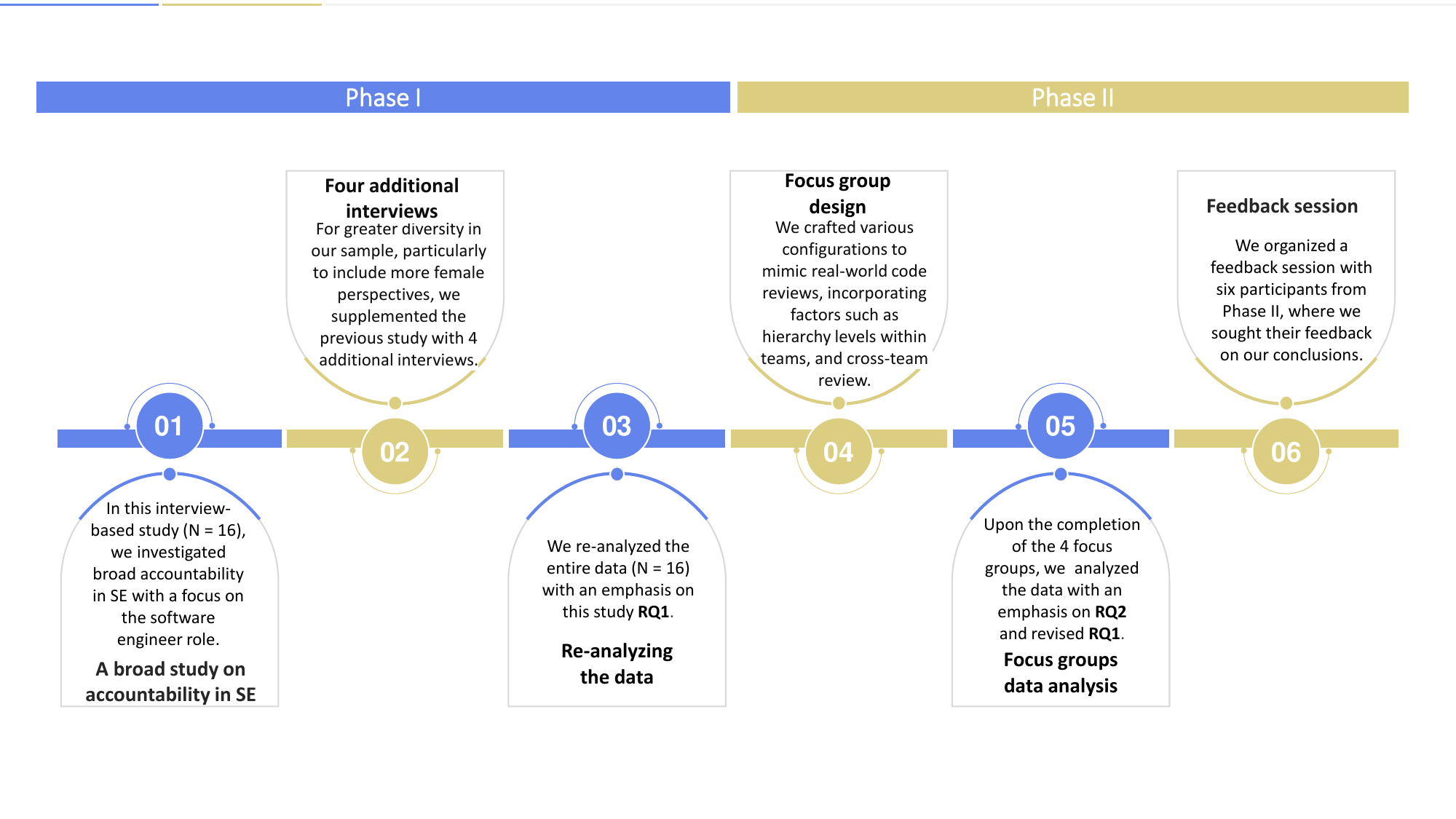}
        \caption{A summary of our research process.}
        \label{fig:research_process}
        \Description[]{Research process flowchart. We begin with the earlier study, then conduct 4 more interviews, re-analyze all the data, design a focus group, analyze the focus groups, and then get feedback from our respondents.}
\end{figure*} 

\paragraph*{Methodological Triangulation}

The use of two methods across sequential phases allowed us to use methodological triangulation \citep{perlesz2003methodological}. By comparing and integrating insights across methods, we refined and contextualized themes, leading to a more informed interpretation. This comparative process enabled us to build a more comprehensive understanding of our themes and the processes we identified. For instance, in Phase I interviews, participants identified pride as an intrinsic driver of accountability, particularly when writing code. However, Phase II focus group discussions revealed a self-regulatory process during collaborative contexts. We learned from Phase II data that individual pride becomes less pronounced to align with collective accountability.

\paragraph*{Integration of Phases I and II} We merged the findings of Phases I and II after Phase II was completed (including data analysis) and during the writing of this manuscript. This part of the investigation is referred to as ``interpretation of the related outcomes'' \citep{creswell2017designing}. Section \ref{sec:findings} presents the results and discusses how far Phase II findings corroborate earlier results in Phase I. We also explore how the outcomes of both stages sync and complement one another.

This integration process entails juxtaposing the findings from Phase I with those from Phase II to identify consistencies, discrepancies, and complementary and new insights. However, in the case of our study, we find mostly complementary and new insights. For example, self-regulation, a new pattern code that has emerged in this exercise, shows that software engineers regulate or tune down some of their intrinsic drivers to accommodate collective consensus and shared accountability. The data collected in the LLM-led reviews also helped us to understand new insights on the impact of AI on accountability.

During this process, we employed a comparative method to analyze how the themes and patterns identified in Phase I were supported, expanded, or challenged by the data from Phase II and whether some of these insights changed the findings. For example, while collective accountability has emerged as findings in Phase I analysis, we only learned that it transpires once the code becomes subject to collective review. This method aligns with Creswell et al. recommendations for mixed-methods research \citep{creswell2017designing}, which emphasize the importance of integrating qualitative data analysis to enhance the depth and breadth of understanding of a particular phenomenon.

To recap this section, figure \ref{fig:research_process} illustrates our research process. We carried out this study in two phases: an interview-based investigation (N = 16) and a focus group study. For Phase I, to enhance the diversity of our sample, we conducted four additional interviews with female participants. The data from these interviews were re-analyzed with a refined focus on \textbf{RQ1}. In Phase II, we designed and conducted focus groups of code reviews. The data from these focus groups were analyzed, emphasizing \textbf{RQ2}. Finally, we conducted a feedback session with six participants from Phase II, which provided further validation of our findings.

\paragraph*{Informed consent} Informed consents from the interviewees and the focus groups' participants were obtained prior to them taking place in accordance with best practices and institutional requirements of the authors' institutions. 

\subsection{Replication package} \label{sec:replication} We share our data and other artifacts at \href{https://doi.org/10.5281/zenodo.14601149}{link.}\footnote{\url{https://doi.org/10.5281/zenodo.14601149}} Interviewees consented to sharing anonymized interview and focus group transcripts.

%% file: findings.tex
\section{Findings}\label{sec:findings}

\begin{figure*}[!t]
    \includegraphics*[trim=2cm 4.5cm 1cm 1cm, clip, width=1.0\textwidth]{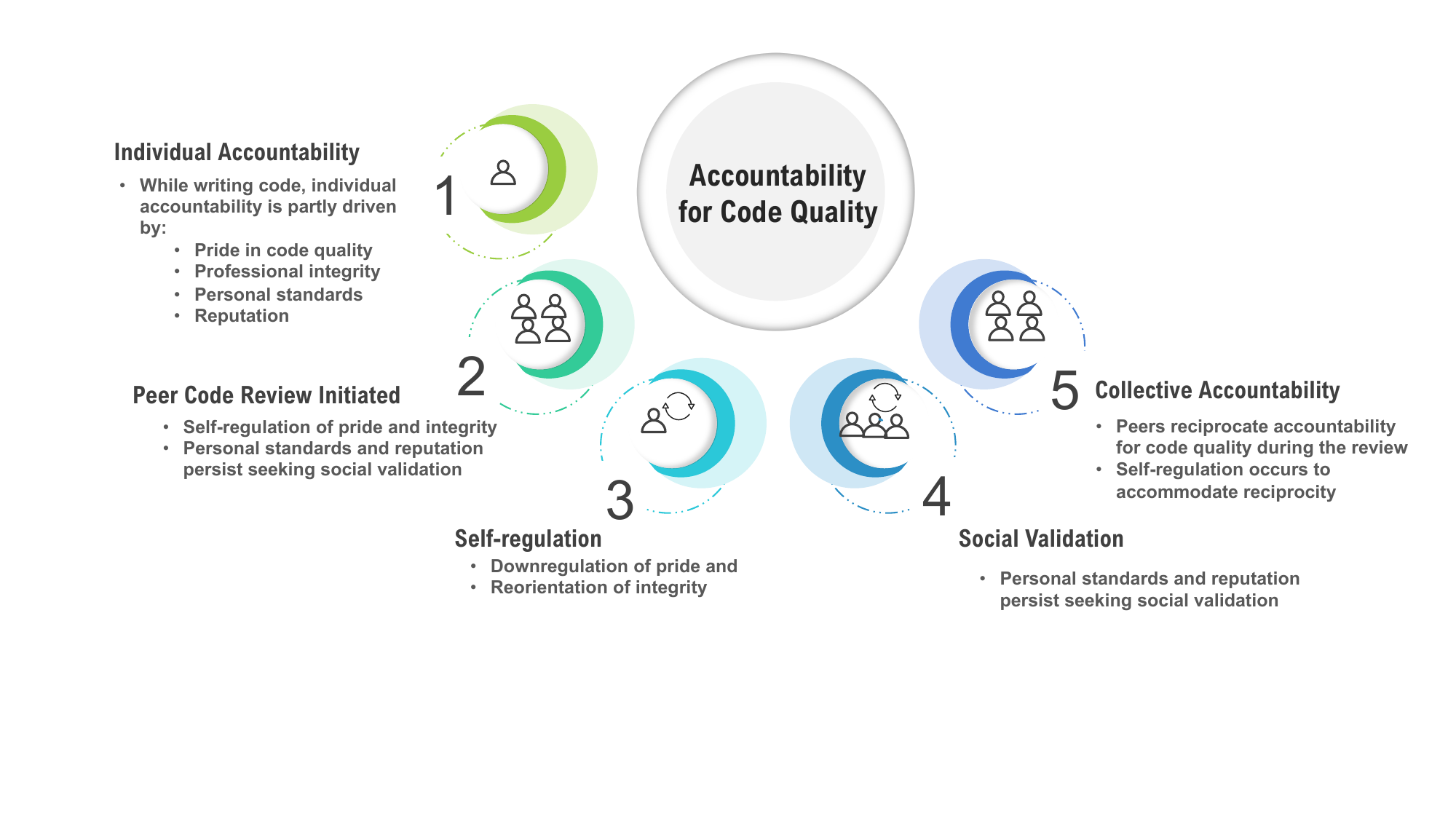}
        \caption{The process of accountability for code quality combining Phase I \& II findings.}
        \label{fig:conceptmap}
        \Description[]{Theory from the research. Sequence of steps: 1, individual accountability, 2, peer code review, 3, self-regulation, 4, social validation, 5, collective accountability.}
\end{figure*}

Our findings reveal a dynamic and evolving concept: accountability in code review. Figure \ref{fig:conceptmap} encapsulates this central idea, illustrating the transitional process from individual to collective accountability as well as the nuanced behaviors that underpin this shift. Accountability has a dual nature. First, it begins as an intrinsic motivation, driven by personal standards, pride in code quality, professional integrity, and reputation. As code review unfolds, these intrinsic drivers are modulated to facilitate collaboration and the collective pursuit for quality, culminating in a shared sense of responsibility for code quality.

The conceptual map captured in Fig. \ref{fig:conceptmap} highlights how accountability evolves from the individual to the collective level through self-regulation and social validation in peer-led reviews. The intricate relationship between individual and collective dimensions of accountability is complex and underpinned with individual and collective behaviors, which are adapted in a shifting social process.

In the remaining of the section, we first present the findings of \textbf{RQ1} across two subsections. In subsection \ref{sec:f1}, we present how individual's intrinsic drivers influence their individual-level sense of accountability for code quality (stage numbered 1 in Fig. \ref{fig:conceptmap}) and in subsection \ref{sec:f2}, we explain the transition we observed in our data from individual to collective levels sense of accountability  (stages numbered 2-5 in Fig. \ref{fig:conceptmap}). Then we present and discuss the results of \textbf{RQ2} with respect to LLMs.

\subsection{RQ1: Intrinsic Drivers of Individual Accountability for Code Quality}\label{sec:f1}

We define the intrinsic drivers of individual accountability for code quality as they emerged from our data:

\textbf{Personal standards:} A self-imposed commitment to surpass organizational or team code quality expectations. These expectations are deeply individualistic and often involve striving for perfection or exceptional outcomes, \emph{``... to me, like I kind of set myself standards ... I want to do it, like the highest quality that whatever I can deliver''} (P8).

\textbf{Professional integrity:} A commitment to doing the ``right thing'' by adhering to agreed-upon standards for code quality, even in the absence of external enforcement. This intrinsic driver captures a sense of responsibility to one's team, organization and future developers, \emph{``... my integrity matters, because I want my code to survive beyond my tenure on that project''} (P15).

\textbf{Pride in code quality:} A personal sense of satisfaction and fulfillment derived from producing high-quality work. Pride motivates engineers, in our sample, to deliver high code quality, \emph{``I believe the primary motivation for high quality code comes from within it's personal satisfaction and pride one feel in their work''} (P14).

\textbf{Professional reputation:} The recognition and respect gained from producing reliable and high-quality work, which is also seen as a strategy for career development. Participants frequently referenced their reputation as a motivator for accountability, \emph{``...  if my code quality is good, my image as a developer is good ... let's say ... you have to cut out of by 20\% of developers in your team, okay. I feel that I should be amongst the good ones that my number doesn't come in top bottom 20\% of the developer in the team''} (P11).

Recall \textbf{RQ1} sought to understand how software engineers' intrinsic drivers influence their sense of accountability for code quality. Our data shows a strong indication that code review is an important accountability mechanism for code quality. P11's statement shows that code review is not merely a formality but an accountability check for code quality: \emph{``... code review is a big thing, if I'm writing some code, and if it is getting passed by single people ... if there are any bugs or any mistakes, I am also \textbf{accountable for those mistakes}. And my senior people will also be \textbf{accountable because they have also passed the code to the repo} and they have not noticed the quality problems''} (P11). 

We found that software engineers' sense of accountability towards code quality is partly driven by four intrinsic factors: \emph{personal standards}, \emph{professional integrity}, \emph{pride in code quality}, and \emph{reputation} (stage numbered 1 in Fig. \ref{fig:conceptmap}). These traits collectively promote a sense of individual accountability for code quality. This sense of accountability is predominately activated when engineers write code.

As illustrated in Fig. \ref{fig:conceptmap}, these intrinsic drivers contribute to the foundation of individual accountability in the broader process of accountability for code quality (stages numbered 1-5 in Fig. \ref{fig:conceptmap}). The early stages of writing software code activate these drivers. They partially drive individual accountability to meet code quality standards. As depicted in Fig. \ref{fig:conceptmap}, once individuals interact with peers and feedback mechanisms during code review, accountability transitions into a collective level.

\news{Personal standards} For many software engineers in our sample, accountability for code quality is driven by self-imposed personal standards for quality and going beyond internally agreed or established standards. P9 demonstrates this behavior, expressing a strong personal commitment to high quality: \emph{``I like to produce some high-quality software engineering ... I'm feeling accountable ... I feel like it's my responsibility to do my best in order to produce something with quality''} (P9). P8, on the other hand, strives for \emph{``perfection''}, demonstrating high personal standards. When she does not meet her own standards, a feeling of guilt arises, underscoring the strong emotional connection that she has with the quality of her code. She stated: \emph{``... for me personally, I always want ... the perfect code ... it feels like you're kind of bad feeling is to be told your work is subpar quality''} (P8). These accounts depict a picture of software engineers who seem to be deeply invested and accountable for the quality of their code. Their accountability appears to be driven by their personal standards.

How personal standards drive the feeling of accountability for code quality is well-established in our interviewees' accounts. For example, when P6, P8, and P14 were asked why they feel accountable for the quality of their code, they respectively answered: \emph{``I just feel like a level of professional accountability. I just want to make sure I'm an admin to myself ... I can live with myself knowing that I did my best''} (P6), \emph{``I want to build like the best quality that possible that whatever they're beyond expectation''} (P8), \emph{``Personal standards I think ... do my job well.''} (P14). 

In conclusion, for the engineers in our sample, this intrinsic quality motivates them to hold themselves accountable, not just to external expectations but to their own self-imposed standards.

\news{Professional integrity} Professional integrity, as it appears in our data, is a commitment to doing the right thing by proactively pursuing and adhering to standards for quality set by the team or the organization. For P7, it is about avoiding the wrong thing. He stated: \emph{``I was thinking of integrity as, in a general sense, not doing something the wrong way. If you're doing something the wrong way, you shouldn't do it at all''} (P7). P7's integrity implies not just completing tasks but completing them correctly, i.e., having integrity. 

P14 linked professional integrity directly with accountability for code quality: \emph{``as a developer, several factors contribute to my sense of accountability for code quality. Professional integrity and a commitment to producing high-quality work are among the key elements that drive my accountability, and the impact of my code on the end user is among the key elements that drive my accountability''} (P14).

Professional integrity also appears to be an ethos, a professional responsibility owed to other developers as well as to oneself. P16 explained: \emph{``... senior management just wants results. They don't care ... but I feel myself accountable ... no one will ask me if the code is not clean. But if I'm not be accountable, then the project will be a mess ... then new developers will not be able to understand anything. And it will be a big problem for the organization. So that's why I am taking care of the accountability''} (P16). Together, these accounts indicate that professional integrity seems to drive software engineers' individual accountability for code quality.

\news{Pride in code quality} While integrity is about doing the right thing, pride is an outcome engineers seek for personal satisfaction achieved from writing quality code. P13 explains: \emph{``so I write code, and I feel like my reward is like internal rewards ... because I produced good code and I'm proud of it ...  if I write good code, then I feel happy, then I should write more good code, because I feel proud and more happy more often''} (P13).

When P7 was asked to explain why he feels accountable for delivering code that ``works properly,'' he explained: \emph{``you have to have some integrity in your work and and pride in your work''} (P7).

These accounts illustrate how our interviewees derive personal satisfaction from their work, which in turn seems to maintain a sense of individual accountability for their code quality.

\news{Professional reputation} Phase I data showed that software engineers perceive their professional reputation as a crucial part of their career, and the quality of their code is tightly linked to their reputation. P14 bluntly explained how her professional reputation is ``closely tied'' to the quality of her code: \emph{``... maintaining a positive professional reputation is closely tied to the quality of the code I produce''} (P14).

P12 eloquently demonstrates this motivation. When she was asked why she ``cares'' about the quality of her code, she replied, \emph{``so I care less about salary and more about the professional image because I think that is what will be useful in the long run and not the salary. Because if we maintain a good image, one day or another, my salary will get up''} (P12). This account shows a strategic approach to career, where the reputation of writing quality code becomes an asset to use in future opportunities. 

It seems that software engineers feel accountable for the quality of their code because they carefully cultivate and protect their reputation through consistent and high-quality code. By writing high-quality code, engineers reputation becomes a reflection of their dedication, skill, and reliability. P15 sums up, when prompted to explain why she feels accountable for meeting high coding standards: \emph{``I don't want to have a bad reputation in the future''} (P15).

This drive to maintain one's professional reputation is tightly linked to a sense of individual accountability for code quality in our data. For example, when P6 was asked why he feels accountable for the quality of his work, he replied: \emph{``if anything goes wrong, it falls on me ... it makes me look bad ... if there's, you know, major bugs found''} (P6). When prompted to explain the reward he received from being accountable, he stated: \emph{``I think just the recognition''} (P6). This account shows the intrinsic link between professional reputation and the sense of individual accountability for code quality amongst the interviewees in our sample. In the case of P6, it seems that his concern to compromise his reputation drives him to adhere to quality standards and take responsibility for his work. The recognition he receives may reinforce his professional reputation. In sum, the pursuit to maintain a strong professional reputation compels engineers, in our sample, to feel accountable for the quality of their code.

\subsection{RQ1: A Transition From Individual to Collective Accountability}\label{sec:f2}

Figure \ref{fig:conceptmap} shows how we conceptualize our key findings. We found a transition in \emph{who} is accountable during code review and the dynamics in the shift from \emph{individual} to \emph{collective} accountability. This transition is a complex interaction of behaviors and collaborative influences. In this section, we unpack these dynamics, grounding each stage of the transition in evidence from our focus group data.

Our analysis shows that ``collective'' is not merely the aggregation of individual contributions in the focus groups but an emergent property of a social process of code review and shared commitments that elevate the sense of accountability from personal achievements to team-based outcomes. In our findings, ``collective'' represents the coalescence of individual behaviors, norms, and responsibilities into a cohesive group dynamic, which we observed in the focus groups.

Individual accountability influenced partly by intrinsic driver (stage 1 in Fig. \ref{fig:conceptmap} numbered 1 at the top left of the figure). As code review begins, pride is downregulated (stages 2 \& 3 in Fig. \ref{fig:conceptmap}, numbered 2 and 3) to allow for seamless integration of feedback and the transition to shared accountability for code quality. P27 explains, \emph{``so when delivery [of feedback] was very good, it's less about pride but more about holding each other accountable for our quality''} (FG2, P27).

Professional integrity is also modulated (stages 2 \& 3 in Fig. \ref{fig:conceptmap}) to accommodate the collaborative efforts toward code quality. It shifts from the individual motivation of doing the right thing to the behavior of avoiding defensiveness and constructively engaging with feedback. P22 explains that receiving feedback during the focus group was not a threat to his professional integrity, \emph{``... when I want to write good code, I try to, and then it turns out that that's actually not so good. And then I get corrected, and I write better code, and I'm better next time. So I don't see it as like challenging my professional integrity ... I see it as an opportunity''} (FG1, P22). This response shows how professional integrity, an individual trait, is regulated during the review process. By framing feedback as an opportunity to improve, P22 demonstrates an alignment of his individual motivation with collective goals, fostering collaborative accountability for code quality.

As the review continues, engineers' personal standards and reputation persist in the review process in \textbf{pursuit of social validation} (stage 4 in Fig. \ref{fig:conceptmap}). Software engineers display their personal standards to their peers, seeking recognition to maintain a reputation within the team. P17 explains, \emph{``... we also like to show and demonstrate our knowledge to other people, that they missed something and we found it... So this also like, I think, improves the impression of other teammates of me as well''} (FG1, P17). P27 explains that code quality is an investment in building his reputation, \emph{`` ... I want to be a better coder or be as good of a coder as I can be. And his [reviewer] feedback helps me with that, you know, which in turn leads to more accountability for my part, you know. A better reputation ... and my code quality got better and being known as a good coder''} (FG2, P27).

This harmonization of intrinsic drivers is to achieve alignment with the team's consensus and expectations, as \textbf{accountability is no longer solely individual but collective} (stage 5 in Fig. \ref{fig:conceptmap}). P20 explains, \emph{``so, I personally see code review as a part of the software quality. So basically, when a code ... is reviewed ... then both the author of that PR and reviewers are like accountable''} (FG1, P20). P25 rationalizes the reciprocity of accountability, when the review takes place, \emph{``I'm doing a review, what this means is ... we need to hold each other accountable and ourselves accountable ... So if I review something and I approve it ... I'm also on the line at that point, not just the author''} (FG2, P25).

In the example of P25's comment, the recognition of peers as key stakeholders in the review process further supports this transition from an individual sense of accountability to a shared one.

The culmination of these dynamics, as depicted in stage 5 of Fig. \ref{fig:conceptmap}, is a shift to collective accountability. This stage is characterized by shared responsibility, where both authors and reviewers internalize accountability for the quality of the code.

Collective accountability, as observed in our data, primarily involves those participating in the code review process, but it also extends in practice to the broader development team collaborating to ensure code quality. We identified this transition in our data further in the linguistic shift among participants, which demonstrates a transition from individual to collective accountability during the review process. For example, participants moved from referencing their accountability individually in the interviews, e.g., \emph{``you know, just accountability towards myself''} (P6) and \emph{``I just care about my work''} (P10), to using inclusive language such as ``we'' and ``each other,'' signifying shared responsibility. One participant explicitly noted, \emph{``we need to hold each other accountable and ourselves accountable''} (FG2, P25). Similarly, another participant emphasized, \emph{``When a code... is reviewed... then both the author of that PR and reviewers are like accountable} (FG1, P20). These shifts in language underscore the evolution of accountability into a collective endeavor fostered in peer reviews.

The sense of accountability among authors is often tied to their individual contributions, emphasizing ownership over their work. For instance, authors expressed pride and satisfaction when producing error-free and high-quality code. Their accountability was heightened by the desire to meet or exceed expectations set by themselves or by their teams. On the other hand, reviewers perceive accountability through the lens of collective responsibility. Reviewers focus on ensuring that the code aligns with team standards and best practices, emphasizing collective ownership of code. Reviewers often exhibit accountability by reciprocating the responsibility shown in the effort of the authors, as demonstrated in the code quality. If not, by providing constructive feedback to improve the code quality. For example, in the quotes shared above, P11 (author) emphasized their accountability as an author, stating, \emph{``... if there are any bugs or mistakes, I am accountable because I wrote the code.} Conversely, P25 (reviewer) highlighted his perspective: \emph{``When I approve a pull request, I am also accountable. If there is an issue later, it reflects on me as well.}

In sum, the focus group data analysis revealed a transition from individual to collective accountability during the code review process. While authoring code, individual accountability for its quality is partly driven by intrinsic factors such as pride, professional integrity, personal standards, and reputation. As code review progresses, these intrinsic drivers are modulated to facilitate feedback integration and collaboration. Subsequently, the focus shifts from individual to shared accountability. During the review, software engineers appear to seek social validation from their peers. This behavior is exhibited through showcasing personal standards in coding, which also serves an investment in maintaining professional reputation. It appears that the process culminates in a collective sense of accountability, where both code authors and reviewers share responsibility for the code quality.

\subsection{RQ2: LLM Disruption to Collective Accountability}

\textbf{RQ2} sought to understand how the introduction of an LLM-assisted review may impact the traditional fabric of peer-led review, especially accountability for code quality. We found that the introduction of an LLM (specifically, a large language model based assistant such as ChatGPT) into the social system of code review causes a disruption to the inherent social dynamics of the process and to the transition of accountability from individual to collective. This disruption is caused by four factors: Absence of \textbf{reciprocity of accountability}, \textbf{human interactions}, \textbf{social validation}, and \textbf{lack of trust in LLM technology}.

The transition to collective accountability (stage 5 in Fig. \ref{fig:conceptmap}) is disrupted by the LLM, because of the \textbf{absence of reciprocity}. Software engineers in our sample described the LLM as a ``machine'' that cannot be held accountable, e.g., \emph{``I would not take its [ChatGPT] suggestions as seriously as a coworker, simply because you cannot hold the model accountable''} (FG2, P25). P26 echoes this view: \emph{``now if LLM is reviewing my code, then I'll be the only person responsible ... you cannot blame or hold the LLM accountable''} (FG2, P26).

This disruption also occurs because software engineers' sense of accountability for their code quality is deeply rooted in the \textbf{human interactions} inherent to code review. For example, P35 perceives human interactions as a fundamental driver of his accountability, and code review is not merely a procedural task, \emph{``the human interaction would actually make me feel more accountable''} (FG4, P35).

LLM-assisted code review also \textbf{challenges the social validation} in the process. Code review is more than a technical or procedural event; its role extends to a process of social validation, where engineers seek peer recognition and personal satisfaction. P36 asserts, \emph{``... there will be less accountability. Because there is no reputation in the game. There is no personal pride in the game. There is absolutely not that... with LLM, there is less accountability''} (FG4, P36). P31 explains that he will miss the personal satisfaction (i.e., ``joy'') he experiences when he receives positive feedback from his peers: \emph{``I wouldn't get the same joy. Because it's just a machine''} (FG3, P31). LLM-assisted code review may challenge the deeply ingrained social validation integral to the traditional peer review process. 

In contrast to an LLM (just a ``machine''), engineers also seek and value the human touch associated with the feedback, \emph{``my concern before reading the review was that it's not a human being ... I probably would still prefer a human interaction''} (FG3, P30). Similar sentiment shared by P31, \emph{``I would just still appreciate the more human at this point ... I could just provide more political review to make the reviewer happy and be kind to them.''} (FG3, P31). This preference shows the inherent human need for social connection and validation, which enhances engagement and trust within collaborative environments. The desire for human interaction reflects a deeper value placed on peers feedback exchanges, which fosters a more human and meaningful review process to software engineers.

Finally, our participants appear \textbf{not to trust in LLM technology} (as of early 2024!). Note that all of our informants reported using LLMs on a daily basis, so unfamiliarity is not the explanation for this. They cited limitations discouraging them from considering it equal to their peers. P27 bluntly asserts: \emph{``It's [ChatGPT] not a real thing ... it has no idea in terms of context what it's predicting ... And I'm not accountable to it''} (FG2, P27). P30 corroborates: \emph{``ChatGPT is not the word of God... it's not something that you should trust''} (FG3, P30). This is mostly due to a perceived lack of shared understanding of the code's context or intent. P20 said: \emph{``some of those [comments] are not like, applicable since it [ChatGPT] does not see the rest of the code, and it doesn't know the full codebase''} (FG1, P20).

The trust issue was persistent in the focus group discussions, despite the overwhelming acknowledgment among the participants of the high quality of the LLM's reviews. During the fourth group discussion, one participant appreciated the LLM highlighting security errors in his code and the overall quality of the LLM's review, \emph{``there were some security feedback, considerations ... in case maybe I didn't think of that it might be something I would think of it now ... I think it [LLM-generated feedback] was generally good''} (FG4, P37). P17 eachoed similar assessment of the LLM's feedback, \emph{``It [LLM feedback] was quite great. I was impressed how, from the code, it connected information and summarize, but the function that's also it gave really good feedback and suggestions on how to improve the code''} (FG1, P17).

The disruption we observed not only impacts the accountability process, but also influences the level of engagement with the LLM-generated feedback in comparison to that of peers. For instance, one participant account indicates selective consideration of the code improvement suggested by the LLM compared to their peers. He stated: \emph{``I think my behavior would be a bit different [for LLM's review] ... I might not take it. Like I might not consider it incorporating ... unless it is a very big issue''} (FG1, P21).

However, our participants praised the educational value of an LLM, and its ability to be leveraged for filtering obvious errors. For example, P31 describes ChatGPT's review of his code as \emph{``educative''} (FG3, P31). Our analysis shows that there is a willingness to use LLM as the first reviewer to filter obvious errors; P18 suggests: \emph{``I would not use something like ChatGPT as the only mechanism for code reviews; I would rather use it as a first level of review before I submit my PR for my peers to review''} (FG1, P18).

%% file: discussion.tex
\section{Discussion and Implications}\label{sec:discussion}

We begin this section with an interpretation of our findings through the lens of self-determination theory (SDT) \citep{deci2000and}, as well as how we contribute to existing accountability theories \cite{frink1998toward,hall2003accountability}. Then, we highlight the implications of our findings on practice and future research directions.

SDT is relevant for the context of our study because it provides a robust reference framework for understanding the intrinsic motivations that underpin software engineers' sense of accountability for code quality. Specifically, SDT's focus on autonomy, competence, and relatedness \citep{deci2000and} resonates with our findings. For instance, the absence of reciprocity in accountability in LLM interactions challenges engineers' sense of relatedness, a core pillar of SDT.

Accountability theories \cite{frink1998toward,hall2003accountability} cover the mechanisms by which individuals feel responsible for their actions within a social and organizational context. This theoretical lens allows us to understand accountability in the context of SE and how the introduction of AI technology alters traditional peer review accountability. Accountability theories provide a reliable foundation to understand the behaviors we observed in this study. This well-established theoretical background is also a good reference to understand to what extent SE aligns or deviates from other contexts.

\paragraph*{Self-determination theory (SDT)} SDT suggests that the pursuit of goals and their attainment is driven by basic psychological needs, competence, relatedness, and autonomy \citep{deci2000and}. These needs are ``innate psychological nutriments that are essential for ongoing psychological growth, integrity, and well-being,'' rather than learned or physiological \citep{deci2000and}.

Competence is a propensity to make an impact on the individual's environment as well as to attain valued outcomes within it \citep{deci2000and}. Intrinsically driven behaviors arise from individuals' desire for competence and the need to be self-determined \citep{deci2000and}. Our findings echo similar principles, demonstrating that software engineers' intrinsic drivers such as professional integrity, personal standards, pride in code quality, and reputation significantly influence their sense of accountability for code quality. This highlights the importance of nurturing the individual sense of competence and autonomy amongst software engineers to foster a stronger accountability for code quality. These findings are also a call to appreciate the psychological underpinnings of accountability for an important outcome like code quality in software engineering. It challenges the assumption that code quality maybe achievable solely by promoting standards, tools, and processes.

Relatedness is the feeling of connection to others, ``to love and care, and to be loved and cared for'' \citep{baum2017optimal,ryan1996supportive}. SDT posits that intrinsic motives flourish in an environment where individuals have a sense of secure relatedness and support \citep{deci2000and}. This explains some of our findings. Software engineers expect their peers to be supportive, empowering them to show greater intrinsic motivation for the accountability of their code quality.

Autonomy ``refers to volition, the organismic desire to self-organize experience and behavior and to have activity be concordant with one's integrated sense of self'' \citep{deci2000and}. SDT suggests that fostering a sense of autonomy enhances intrinsic motivation and leads to better performance and well-being \citep{reeve1998autonomy,deci2000and}. By cultivating an environment that supports competence, relatedness, and autonomy, SE teams and organizations may enhance the intrinsic drivers of software engineers, leading to a heightened sense of accountability for code quality. 

These findings may indicate that the current focus on external controls (e.g., quality assurance practices) and incentives (e.g., promotion), as exemplified by industry-oriented metrics approaches such as DORA~\cite{forsgren2018} need to be combined with socially-embedded antecedents to drive code quality. For example, Alami and colleagues reported that psychological safety, or team level perception of non-judgmental interactions, openness and emotional security, also enhances team's ability to pursue quality expectations \citep{alami2024role}.

Although SDT claims that intrinsic motivations are goal-directed behavior either in an autonomous or controlled environment \citep{deci2000and}, our work shows that is not always the case. Intrinsic drivers can also manifest for an outcome that is not necessarily tangible yet, such as feeling and showing accountability for code quality. In our work, the ultimate outcome would be better code quality; however, we have only demonstrated evidence for behaviors showing accountability for this outcome. Hence, we contribute to SDT by highlighting the nuanced ways intrinsic motivation can drive behaviors focused on accountability, even when the end goal is not immediately realized.

\paragraph*{Accountability theory} The conceptualization of accountability captures both the formal and the informal manifestation of accountability \cite{frink1998toward}. While informal accountability is formalized by institutionalized rules and policies \cite{frink2008meso}, informal (grassroots accountability or accountability to peers \cite{alami2024understanding}) uses rules and norms outside the formal organizational context \cite{zellars2011accountability}. Informal accountability is grounded in unofficial expectations and discretionary behaviors that result from the socialization of network members \cite{romzek2012preliminary}. Shared norms also lay out an informal code of conduct used by group members as a reference for appropriate and inappropriate behaviors \cite{romzek2012preliminary}. Romzek et al. found that informal accountability in nonprofit networks is fostered by trust, reciprocity, and respect for institutional turf \cite{romzek2012preliminary}. Similarly, informal accountability is exercised through evaluations that result in either rewards or sanctions \cite{romzek2014informal}, but remain informal in nature. For example, rewards can be in the form of favors and public recognition, and sanctions may lead to reduced reputation, loss of opportunities within the group, and exclusion from future information sharing \cite{romzek2014informal}.

Our study contributes to existing accountability theories by showing that felt or individual accountability is temporal in the context of teamwork. In the context of SE, this individual accountability persists while writing code and prior to the review, then shifts to a collective level to become a shared accountability. This temporality is interrupted when an accountability mechanism such as code review takes place. We also contribute to this theoretical landscape by identifying some individual accountability antecedents. Hall et al. state that ``relatively little empirical work is available to inform our perspectives of antecedents to accountability. Many constructs that would seem to be antecedents to felt accountability'' \citep{hall2017accountability}. In the context of SE, our findings suggest that intrinsic drivers like professional integrity, personal standards, pride in code quality, and reputation serve as key antecedents to individual accountability. This understanding can inform the design of more nuanced accountability mechanisms that capitalize on personal motivations, ultimately improving individual and collective levels of accountability for code quality.

Our research also synergizes SDT and accountability theories by elucidating that intrinsic drivers linked to the psychological needs outlined in SDT---competence, relatedness, and autonomy \citep{deci2000and}---play a role in fostering accountability among software engineers. This connection suggests that nurturing these intrinsic drivers may lead to a stronger individual and collective sense of accountability for code quality. Our findings imply that integrating SDT principles into team-level accountability frameworks and mechanisms may promote more effective and psychologically supportive environments conducive to enhancing accountability for code quality.

\subsection{Industry Implications}

\noindent For implications on practice, we discuss four key takeaways from this study and their implications: (1) code quality through accountability, (2) promoting collective code Ownership, (3) Aligning SE education with its social dynamics, and (4) integration strategies for LLMs.

\paragraph*{Code Quality Through Accountability}

\noindent Even though code review has been extensively investigated \citep{davila2021systematic,badampudi2023modern}, the power of social dynamics inherent in the process to foster accountability for code quality has been underestimated. Our study shows that ensuring code quality extends beyond a technical endeavor. The pursuit of code quality is also embodied in a deep sense of accountability in software engineers' work. While engineers' intrinsic drivers influence their accountability, the shift to shared accountability during the review marks a departure from viewing code quality as a personal and technical achievement to a collective endeavor.

Code quality is not merely a set of standards or metrics but a shared value that is cultivated through a sense of individual and collective accountability. Alami and Krancher drew similar conclusions \citep{alami2022scrum}. They found that Scrum's social practices foster ``social antecedents'' (e.g., psychological safety and transparency) conducive to cultivating behavior and commitment to software quality \citep{alami2022scrum}. Amongst these ``social antecedents'', a sense of collective accountability for a team's outcomes, including software quality, is more pronounced amongst developers.

A useful way to think about code review as a social system is developed in \emph{social learning theory}~\citep{bandura1977social}. Social learning theory (SLT) posits that individuals learn new behaviors and norms through observing and imitating others \citep{bandura1977social}. In order for this social learning process to take place, Fogarty and Dirsmith suggest that socialization practices such as mentoring facilitate ``normative'' and ``mimetic'' isomorphism, especially for standards and behaviors established and/or sanctioned by a profession \citep{fogarty2001organizational}. New members imitate their mentor's performance in their roles to fit within the team and the organization and embody the profession's ethos. They also actively adopt new skills and mimic their mentor's behaviors and values to advance their career in the organization \citep{fogarty2001organizational}.

SLT is relevant in understanding the role of personal qualities intrinsic to a software engineer, such as pride in code quality, professional integrity, upholding personal standards, and maintaining a reputation. The traditional mechanisms of social learning rely primarily on observation, imitation, and social interaction. During mentoring, not only are technical skills passed on but also the ethos of the profession, including values such as pride in code quality, professional integrity, and individual accountability for code quality.

\begin{tcolorbox}

\textbf{Mentoring}: To leverage the potential of social learning and promote accountability for code quality, organizations could implement structured mentorship programs. Experienced engineers with strong intrinsic accountability could be paired with new members. In this socialization process, mentors become role models, and reciprocal determinism may take place.

\end{tcolorbox}

Incorporating structured mentorship programs could promote and embed intrinsic drivers of accountability within the organizational culture by emphasizing the professional values and social norms that underpin code quality. These programs would involve pairing experienced engineers with newcomers to foster a culture of accountability for quality. Mentors with strong intrinsic accountability could model behaviors such as professional integrity, pride in work, and professional reputation. Through ongoing mentoring and feedback, mentees may internalize these values, which may align newcomer behaviors and lead to long-term behavioral shifts \citep{ragins2007handbook,johnson2004mentoring}.

\begin{tcolorbox}

\textbf{Mentorship programs to promote accountability}: Structured mentorship programs could include the following components:\\

- Goal-oriented pairing: Pair new engineers with mentors based on shared professional goals, technical expertise, or accountability practices \citep{ragins2007handbook}.\\

- Regular feedback sessions: Create opportunities for mentors to provide constructive feedback and discuss the ethos of professional integrity, pride in code, and maintaining quality standards \citep{ragins2007handbook,london2002feedback}.\\

- Accountability exercises: Include activities such as collaborative code reviews, where mentors model behaviors like ownership of code quality and providing actionable feedback \citep{london2002feedback}.\\

- Career development integration: Tie mentorship outcomes to career progression to reinforce the mentee's commitment to accountability as a professional standard \citep{allen2004career}.\\

- Reciprocal learning opportunities: Encourage mentees to share their own perspectives and practices, fostering mutual growth and innovation \citep{johnson2004mentoring}.\\

\end{tcolorbox}

The transition to collective ownership and accountability for code quality during the review process highlights the need for constructive delivery of the feedback to foster this communal approach, especially given that the process is susceptible to perceptions of unfairness \citep{german2018my} and negative impressions when the code is substandard \citep{bosu2013impact,bosu2016process}. Feedback intervention theory (FIT) suggests that feedback is effective when its focus is the task and not the individual, the intent is to facilitate learning, and it is perceived as relevant by the recipient \citep{kluger1996effects}. Our findings underscore the importance of the recipient's openness to feedback and the reviewer's constructive delivery in order for the improvement to materialize.

The process where individual accountability for code quality shifts to collective is contingent on this constructive focus. Some of our participants deliberately ``read the room'' or ``politically'' dress their feedback with ``kindness'' and an intention to make the author of the code happy. An average computer science graduate may not be  equipped with the skills to handle this delicate process. The lack of, and need for these skills was recently acknowledged in the ACM 2023 curriculum revision, i.e.,  ``[m]ore focus on team participation, communication, and collaboration''~\citep{ACMed23}.

\begin{tcolorbox}

\textbf{Feedback:} Organizations should develop and continuously improve guidelines for giving and receiving feedback that emphasize constructiveness, learning, and improvement. Educators should emphasize these skills in educational settings.

\end{tcolorbox}

\paragraph*{Promoting Collective Code Ownership}

The shift from individual to collective accountability emphasizes the principles of collective ownership of code \citep{greiler2015code,bird2011don}. When fostering a culture of collective ownership of code and its quality, accountability for quality transcends individual levels to become a shared group ethos. Social identity theory posits that when individuals see themselves as part of a collective entity, they share the responsibility for the group's success. \citep{tajfel1979integrative}. The identification with the group enhances motivation to contribute, as team members derive intrinsic satisfaction from the group's achievements \citep{tajfel1979integrative}.

In SE, research shows that individual silos can create barriers to knowledge sharing and collaboration, while collective ownership promotes a shared responsibility for the codebase \citep{bird2011don}, reinforcing the importance of shared accountability in the pursuit of software quality \citep{alami2024understanding}. Modern code review practices also demonstrate that when multiple team members contribute to the review, the process enhances code quality by broadening the accountability landscape and strengthening the sense of shared responsibility for the codebase at the group level \citep{greiler2015code}. Thongtanunam and Tantithamthavorn reported that diverse metrics of ownership, such as commit- and line-based measures, accommodate varied contributions, fostering team cohesion and a sense of shared purpose \citep{thongtanunam2024code}. These findings align with the principles of social identity theory that collective ownership enhances team identification and intrinsic motivation, promoting a group ethos centered on shared success \citep{tajfel1979integrative}. Supported by previous work and social identity theory, our findings indicate that code review should remain a core SE quality practice.

\begin{tcolorbox} 

\textbf{Promoting collective code ownership practices:} Organizations should establish processes that encourage shared responsibility for the codebase. Our findings show that code review serves as a fundamental mechanism for promoting collective ownership principles in software engineering workflows. These findings suggest that maintaining and enhancing code review practices can reinforce team collaboration and intrinsic motivation, driving higher accountability for code quality and a stronger group ethos.

\end{tcolorbox}

Educational programs should align with the nature of SE practices in the industry, as previously mentioned. Our findings demonstrate that SE is a socially loaded practice. Future software engineers should be prepared to thrive in such environments. This aspect of SE was highlighted as far back as 1988 in Curtis et al. \citep{curtis1988field} and re-emphasized repeatedly in many studies, e.g., \citep{demarco2013peopleware,salas2008teams,begel2008pair,ACMed23}.

\begin{tcolorbox} 

\textbf{Aligning SE education with its social dynamics:} To support collective accountability for code quality in SE, educational programs should train software engineers in collaborative practices, constructive feedback techniques, and interpersonal communication. These skills are critical for fostering collective ownership in team-based environments.

\end{tcolorbox}

\paragraph*{Integration Strategies for LLM}

The transition to collective ownership and accountability for code quality during the review process highlights the need for constructive delivery of the feedback to foster this communal approach, especially given that the process is susceptible to perceptions of unfairness \citep{german2018my} and negative impressions when the code is substandard \citep{bosu2013impact,bosu2016process}. Feedback intervention theory (FIT) suggests that feedback is effective when its focus is the task and not the individual, the intent is to facilitate learning, and it is perceived as relevant by the recipient \citep{kluger1996effects}. Our findings underscore the importance of the recipient's openness to feedback and the reviewer's constructive delivery in order for the improvement to materialize.

\begin{tcolorbox}

\textbf{Complementary role, not a replacement:} To retain the social-centric aspects of accountability for code quality and the social validation process built in, LLMs should be integrated as aids to human expertise rather than a replacement.

\noindent \textbf{LLM as first-line reviewer:} One potential integration is to deploy an LLM as a preliminary reviewer to filter out straightforward issues before the actual peer review takes place. This implementation will also permit leveraging LLMs in an educational role.

\end{tcolorbox}

While LLMs may bring technical and educational augmentation to the process, as seen with chatbots~\citep{Wessel2022}, they also impact the social learning process. It is a considerable shift from the traditional peer-led review, with the potential to also disrupt well-established patterns of social learning and accountability. Their inability to simply be human, reciprocate accountability, and participate in social interactions hinders the social learning framework that underpins team dynamics and collective accountability for code quality. The effectiveness of social learning, as posited by SLT \citep{bandura1977social,bandura1986social}, relies on the model role, in which skills, behavior, norms, and values are adopted during a complex normative adjustment. The assumption that the integration of AI into SE is a matter of plug and unplug is naive.

\subsection{Research Implications}

Our study extends accountability theory by demonstrating the temporal and dynamic nature of individual and collective accountability in SE. The shift from individual to collective accountability, mediated by intrinsic drivers such as professional integrity, pride in code quality, and personal standards, demonstrates the significance of informal and nuanced aspects of accountability beyond formal organizational structures. These findings open avenues for further research to explore the tensions and synergies between institutionalized and informal accountability mechanisms. For example, in our previous work, we reported performance reviews as a formal accountability mechanism \citep{alami2024understanding}. For instance, how do formal mechanisms like performance reviews interact with informal accountability practices, such as peer validation or social learning, to shape engineers' sense of accountability? Additionally, which type of accountability mechanism---institutionalized or informal---exerts a more significant influence on fostering accountability for code quality, and under what circumstances?

The temporal aspect of accountability, we found in this study, enriches existing frameworks, providing empirical evidence for how individual perceptions and intrinsic motivations can evolve into shared accountability within a collaborative process like code review. By identifying intrinsic drivers as antecedents to accountability, our findings address the noted gaps in accountability theory, as highlighted by Hall et al. 2017 \citep{hall2017accountability}, and offer a pathway for designing accountability mechanisms that leverage personal motivations in SE and similar socio-technical domains. However, we still do not know how to operationalize these personal qualities into concrete practices to harness their power. For instance, future research should explore this broad question: What are the mechanisms through which intrinsic motivations can be integrated into formal and informal accountability structures to enhance individual and collective accountability in SE and similar socio-technical practices?

The integration of AI tools like LLMs into SE processes challenges traditional accountability mechanisms and social dynamics, necessitating a thoughtful design approach. Our findings emphasize the need to preserve the social integrity of SE practices by ensuring that AI complements rather than replaces human interactions. For instance, deploying AI-led reviews as first-line reviewers can streamline technical assessments while leaving the social and collaborative dimensions of accountability intact for peer review. Additionally, the educational role of LLMs offers potential for skill enhancement, but their inability to reciprocate accountability or engage in social validation underscores the importance of maintaining human oversight. These insights call for future research to explore how AI can be integrated into SE in ways that sustain the collaborative and social fabric foundational to the discipline. For example, future work can explore questions like: How can AI tools be designed and integrated into software engineering to preserve the social integrity of its practices?

%% file: validity.tex
\section{Research Trustworthiness, Limitations and Trade-offs}\label{sec:trust}

\subsection{Trustworthiness}

\noindent We implemented several techniques to address the requirements of research trustworthiness \citep{miles2014qualitative}. We reported \textit{Saturation} (Phase I), \textit{Member checking} (Phase I), and \emph{Feedback session} (Phase II) in Sect. \ref{sec:methods}.

\textit{Triangulation}: We triangulated data sources, including interviews, focus groups, and participant feedback sessions. This exercise allowed us to ensure that our findings are corroborated across different data sources and contexts.

\textit{Peer debriefing}: Although the analysis was primarily conducted by the first author, the second and third authors reviewed the proposed codes, and the results were continuously discussed and scrutinized by the other two authors in several meetings throughout the analysis process. The participation of two authors in the coding process helped minimize researcher biases \citep{miles2014qualitative}. This approach is grounded in our epistemological stance, constructivism, which posits that knowledge is socially constructed and that collective intellectual engagement can lead to more reliable understandings of the data \citep{fosnot2013constructivism}. 

\textit{Thick description}: We endeavored to provide a detailed explanation of our research process and the decisions we have made throughout (see Sect. \ref{sec:methods}). In addition, we assembled a comprehensive replication package (see Sect. \ref{sec:replication}).

\subsection{Limitations and Trade-offs}\label{sec:limit}

\textit{Homogeneous sample}: Our sample is composed only of software engineers. In line with roles theory \citep{katz1978social,frink2004advancing}, we limited our sample to the software engineer role to mitigate the potential for variations that may arise by the inclusion of multiple roles. Roles theory suggests that individuals' accountability is closely linked to roles \citep{katz1978social,frink2004advancing}. This narrow focus strengthens the internal validity of our study and allows for role-centric conclusions.

\textit{Focus on intrinsic drivers}: By focusing primarily on intrinsic drivers and their influence on accountability, we may have inadvertently undermined other factors. For example, in our previous work, we identified institutional factors, such as financial incentives or denial of promotions, that also promote accountability in SE environments \citep{alami2024understanding}. 

\textit{Limited variation in the focus group design}: Another tradeoff is the limited number of variations in the focus group configurations, and the code snippets we used were not of industrial caliber. The consistency across the four groups, shown in the collected data and findings, suggests that additional configurations might not have significantly altered the results. In addition, we prioritized in-depth discussions, which may have been diluted by overly complicated configurations and complex code.

Another tradeoff for this study design is with more realistic, complex, and context-aware code. However, we felt this would greatly limit the accessibility of the focus groups. A future study, examining the contextual intricacies of a proprietary codebase, would shed insight on the role of context in this setting.

We conducted focus groups synchronously and online. Often code reviews, in particular on GitHub and similar sites, are asynchronous and text-based. Open source projects have different dynamics than the ones we discuss here. Hence, our findings our findings may not fully transferable to asynchronous or open-source code reviews. Furthermore, the online setting may have influenced participants' behavior differently than an in-person setup. 

The implementation of a pre-focus group questionnaire to mitigate the risk of social desirability bias \citep{furnham1986response} and self-censorship \citep{yanos2008false} during the focus group discussions carries the risk of priming participants. To mitigate this risk, we avoided the explicit use of the word ``accountability'' in the questions. In addition, during the discussions, we asked participants to provide concrete examples to anchor their responses in their personal experiences, thereby avoiding generic or socially desired answers.

%% file: conclusion.tex
\section{Conclusion}\label{sec:conclusion}

\noindent In this study, motivated by our prior work on the broader concept of accountability in SE \cite{alami2024understanding}, we sought to understand the interplay between intrinsic drivers and software engineers' sense of accountability for code quality (\textbf{RQ1}). Also motivated by the evolving nature of AI in SE \cite{fan2023large}, we investigated the impact of the introduction of LLM-assisted review in the context of code review.

We provide insights into how intrinsic drivers, namely professional integrity, pride, personal standards, and reputation, shape engineers' individual accountability. The study also uncovered a complex accountability process that transitions from \emph{individual} to \emph{collective} throughout the process. Finally, the integration of an LLM into this socially loaded process unveiled a pronounced reluctance among software engineers to compromise the social integrity inherent in traditional code review. We contribute to the ongoing efforts to make SE a socially aware practice. The study also created awareness about the integration of AI in SE. The LLM disruption, we observed, highlights emphasizing the preservation of essential human and social aspects such as accountability, social validation, and intrinsic motivation.

Our findings highlight the importance of aligning AI integration with the social dynamics of SE processes to maintain their collaborative essence. Our work opens avenues for future research to investigate mitigating the impact of AI integration in the social dynamics of SE. Future work should explore design frameworks and practical strategies that bridge the gap between the social dynamics inherent in SE and AI integration. Research efforts should ensure that AI augments SE rather than disrupts its critical human-centric practices like accountability and social validation. Such efforts can guide the development of AI that complements the collaborative fabric of SE, fostering both technological efficiency and social cohesion.

For instance, some future research questions to pursue include: How can AI tool integration be designed to preserve the social integrity of software engineering environments and practices? What mechanisms can effectively integrate LLM-generated feedback into socio-technical SE practices without diminishing the human element? Addressing such questions can guide the development of AI that complements the collaborative and social fabric of SE.